# Microtubes and nanomembranes by ion-beam-induced exfoliation of $\beta$-Ga$_2$O$_3$

D. M. Esteves[1,2,*], R. He[3], C. Bazioti[4], S. Magalhães[2,5], M. C. Sequeira[6], L. F. Santos[7], A. Azarov[4], A. Kuznetsov[4], F. Djurabekova[3], K. Lorenz[1,2,5], M. Peres[1,2,5]

[1] INESC Microsystems and Nanotechnology, Rua Alves Redol 9, Lisboa 1000-029, Portugal

[2] Institute for Plasmas and Nuclear Fusion, Instituto Superior Técnico, University of Lisbon, Av. Rovisco Pais 1, Lisboa 1049-001, Portugal

[3] Department of Physics, University of Helsinki, P.O. Box 43, FI-00014, Helsinki, Finland

[4] Department of Physics and Centre for Materials Science and Nanotechnology, University of Oslo, PO Box 1048 Blindern, N-0316 Oslo, Norway

[5] Department of Nuclear Science and Engineering, Instituto Superior Técnico, University of Lisbon, Estrada Nacional 10, km 139.7, Bobadela 2695-066, Portugal

[6] Helmholtz-Zentrum Dresden-Rossendorf, Bautzner Landstraße 400, 01328 Dresden, Germany

[7] Centre for Structural Chemistry, Institute of Molecular Sciences and Department of Chemical Engineering, Instituto Superior Técnico, University of Lisbon, Av. Rovisco Pais 1, Lisbon 1049-001, Portugal

* Corresponding author: duarte.esteves@tecnico.ulisboa.pt

**ABSTRACT:**

This paper reports an innovative process to fabricate $\beta$-Ga$_2$O$_3$ microtubes and nanomembranes based on ion implantation in (100)-oriented single-crystals. We show that, under specific flux and fluence conditions, the irradiation-induced strain profile promotes the detachment and rolling-up of a thin surface layer, forming a microtube. The strain-disorder interplay was investigated in detail for Cr-implanted $\beta$-Ga$_2$O$_3$ with a range of complementary methods, showing an excellent agreement between experimental and simulation data, and suggesting an exfoliation mechanism that is correlated with the anisotropic nature of the $\beta$-Ga$_2$O$_3$ monoclinic system and its easy-cleavage planes. Moreover, these microtubes can be unrolled upon a subsequent annealing step, resulting in nanomembranes with bulk-like crystalline quality that can be transferred to other substrates. The recovery of the implantation-induced damage under thermal annealing has also been studied, showing a remarkable recovery at moderate temperatures (~500 °C). This observation underscores the potential of this method for the scalable production of nanomembranes with improved reproducibility compared to conventional mechanical exfoliation techniques. Importantly, such exfoliation can be done employing different ions, providing simultaneous $\beta$-Ga$_2$O$_3$ doping, chosen to control the structural, optical, magnetic and electrical properties of the nanomembranes, thus tailoring them to fit the desired applications.

## 1. Introduction

As a member of the family of ultra-wide bandgap semiconductors, Ga$_2$O$_3$ has recently attracted significant interest due to its exceptional properties. In its monoclinic $\beta$ phase, which is chemically and thermodynamically stable at the usual device performance conditions, this material exhibits a bandgap of ~4.9 eV at room temperature (RT) and a large breakdown electric field of ~8 MV/cm, thus rendering its Baliga figure of merit larger than those of GaN or SiC [1]. As such, $\beta$-Ga$_2$O$_3$ has been deemed interesting for applications including optoelectronic devices [2], high-power electronics [3], solar-blind photodetectors [4–6], and gas sensors [7], among others. However, in spite of these advantages, $\beta$-Ga$_2$O$_3$ faces challenges such as the lack of reproducible $p$-type doping and the low thermal conductivity. These are properties that hamper the widespread usage of this material [8–10], particularly in optoelectronics and power electronics, where efficient complementary doping and heat dissipation are essential for device stability and performance.

Moreover, in the context of photonic applications, $\beta$-Ga$_2$O$_3$ is a great host material for optically active centers operating over a broad spectral region, spanning from the ultraviolet (UV) to the infrared (IR) [11]. For example, the efficient red/near-infrared emission due to Cr$^{3+}$ intraionic transitions (~650–850 nm) shows great potential for optical detectors of ionizing radiation [12–15]. In particular, this Cr$^{3+}$ emission lies within the so-called first biological window (~700–950 nm), where biological tissue absorbs the least [16], making it specially compelling for in vivo applications. Recent works based on this dopant include tunable optical microcavities [17], non-contact thermometers [18], active and passive dosimeters of ionizing radiation based either on optical (including photo-, thermo- and/or ion-beam-induced luminescence [19–21]) or electrical signals [22].

Importantly, the anisotropic system of $\beta$-Ga$_2$O$_3$ lends itself to two easy-cleavage planes, (100) and (001), in spite of not being a true Van der Waals 2D material [23]. It is thus possible to easily produce thin flakes by the conventional mechanical exfoliation processes (e.g., the scotch tape method). Several works based on these flakes have demonstrated the potential of these structures in different applications [24–26]. However, these conventional techniques often offer limited reproducibility and control over the morphology, thickness and lateral dimensions of the flakes.

In this work, we present and investigate an alternative, recently-patented [27] ion-implantation-induced exfoliation method, with possible advantages in terms of reproducibility, scalability and tunability of the optical, electrical, magnetic or structural properties of the obtained membranes. We show that, upon reaching a disorder threshold, the induced strain and stress fields promote the detachment and rolling-up of a thin surface layer, creating a microtube. By employing a wide range of complementary methods, which showed excellent agreement between experimental and simulation data, we correlate the exfoliation mechanism with the strain and stress profiles induced by implantation defects, in conjunction with the anisotropic nature of the monoclinic system. Spectacularly, these microtubes can be unrolled by subsequent annealing, resulting in nanomembranes with bulk-crystal-like quality transferable to other substrates, such as Si. Moreover, our novel method provides multiple advantages compared with techniques such as the well-known SmartCut® process [28], which involves implanting gas species, such as H or He, into a sample followed by annealing. This process promotes the expansion of the gas bubbles generated during the implantation, causing highly uniform layers to split, yielding wafers with excellent crystalline quality and sharp interfaces. While this is effective for silicon-on-insulator wafer fabrication [28,29], only initial tests have been performed on Ga2O3, leading to rough surfaces requiring additional etching [30]. Furthermore, our approach to obtain nanomembranes out of single-crystal (100)-oriented $\beta$-Ga$_2$O$_3$ wafers offers the integration of additional functionalities, such as tailored optical, electrical, and magnetic properties, directly during the process.

## 2. Experimental and computational methods

For this work, nominally undoped commercial $\beta$-Ga$_2$O$_3$ single-crystals from Novel Crystal Technology, Inc. were used. These crystals were grown by the edge-defined film-fed growth (EFG) method and mechanically cleaved along the {100} easy-cleavage planes to a thickness of ~500 μm and cut to lateral dimensions of 5 mm × 5 mm.

The ion implantations with 250 keV Cr were performed at room temperature at the 500 kV implanter of the Ion Beam Centre of the Helmholtz Zentrum Dresden-Rosendorf (IBC-HZDR) and at the 210 kV high flux ion implanter of the Laboratory of Accelerators of Instituto Superior Técnico, Universidade de Lisboa (IST) [31], with a tilt angle of 7°, to fluences between $6.0\times10^{12}$ and $6.0\times10^{14}$ cm$^{-2}$. Although the equipment at IST is a high flux ion implanter, the results of the implantations in the two laboratories are similar, provided that the implantation fluxes are low enough ($\lesssim 10^{12}$ cm$^{-2}$ s$^{-1}$). The samples were annealed in an N$_2$ atmosphere using a Rapid Thermal Processor AS-One 100.

Stopping and Range of Ions in Matter (SRIM) Monte Carlo simulations [32] were performed in the detailed calculation mode, with full damage cascades, considering displacement energies of 28 keV for Ga and 14 keV for O [33] and a density of 5.88 g/cm$^3$ for $\beta$-Ga$_2$O$_3$ [8]

The Rutherford Backscattering Spectrometry in Channeling mode measurements (RBS/C) were performed in a 2.5 MV Van de Graaff accelerator, using a 2 MeV He$^+$ ion beam with a nominal current of ~4 nA, with a Si PIN diode detector placed at an angle of 140° with respect to the beam direction. The random spectra were obtained by tilting the sample by 5° and rotating the azimuthal angle. The relative defect concentration profiles were extracted by applying the two-beam approximation, using the Defect Concentration (DECO) software developed at the University of Jena [34,35]. As the name suggests, this model divides the analyzing beam particles into two portions: the random fraction, which interacts with all the lattice atoms as in an amorphous target, and an aligned fraction, which interacts solely with lattice atoms that are displaced from their equilibrium positions, be it due to the presence of defects or due to their thermal motion [34–36]. In the framework of this model, the state of a particle can change from aligned to random due to dechanneling, and the probability of such can be analytically calculated [37]. However, the model does not properly account for more complex defects, such as stacking faults, which contribute to dechanneling in a different manner; this was accounted for in this paper by decreasing the critical angle of channeling (i.e., the minimum angle for which the impinging particle leaves the channel). This procedure has been previously shown to yield a good agreement with state-of-the-art RBS/C analysis using Monte Carlo codes in the case of GaN [38]. In the particular case of (010)-oriented $\beta$-$Ga_2O_3$, DECO has been previously used with critical angle pre-factors between 0.65–0.71, which are only slightly lower than the theoretical value of $\sqrt{2}/2$ [39]. The critical angle pre-factors used in the present work were smaller than those reported in ref. [39], of the order of 0.17–0.20. The literature is scarce in the particular case of applying the two-beam model to (100)-oriented $\beta$-$Ga_2O_3$ [40,41], but yielded good agreement between experimental results and SRIM simulations.

The High-resolution X-ray Diffraction measurements (HRXRD), namely $2\theta - \omega$ scans and reciprocal space maps (RSM), were performed in a Bruker D8 Discover diffractometer. The primary beam optics consists of a Göbel mirror, a 0.2 mm collimation slit, and a 2-bounce (220)-Ge monochromator, selecting the copper (Cu) K$\alpha_1$ X-ray line (wavelength of 1.5406 Å). The secondary beam path consisted of a 0.1 mm slit and a scintillation detector. The experimental data were fitted considering the dynamical theory of X-ray diffraction using the Multiple Reflection Optimization package for X-ray diffraction (MROX) 2.0 [42,43]. Considering the measurement of RSM, it is important to emphasize that they can only be acquired in an absolute scale if a reference (i.e., a fixed point) is obtained. In this work, this was achieved employing Bond's method [44] (see the Supporting Information for further details) to measure the lattice parameters of the virgin sample with high accuracy.

Raman data (available in the Supplementary Information) were obtained at room temperature with a LabRAM HR 800 Evolution spectrometer (Horiba, Jobin-Yvon) in a backscattering configuration with an Olympus BXFM confocal microscope with a 100× objective, using a 532 nm laser and a 600 lines/mm grating, with a power of about 10 mW at the samples.

The Scanning Electron Microscopy (SEM) inspections were performed at the RAITH 150 electron beam lithography system, using an acceleration voltage of 10 kV and an aperture size of 15 μm. The images were obtained in cross-section, with the sample surface parallel to the electron beam, in order to assess the thickness of the fabricated microtubes.

Nano-scale investigations were performed by employing (scanning) transmission electron microscopy [(S)TEM] combined with energy-dispersive X-ray spectroscopy (EDX). The experiments were conducted on an FEI Titan G2 60–300 kV microscope, equipped with a CEOS DCOR probe-corrector, monochromator and Super-X EDX detectors. Observations were performed at 300 kV with a probe convergence angle of 24 mrad. The camera length was set to 60 mm and simultaneous STEM imaging was conducted with 3 detectors: high-angle annular dark field (HAADF) (collection angles 101.7–200 mrad), ADF (collection angles 22.4–101.7 mrad) and annular bright field (ABF) (collection angles 8.5–22.4 mrad). The resulting spatial resolution achieved was approximately 0.08 nm. Part of the rolled microtubes were deposited on TEM Cu-grids for observations along the direction perpendicular to the (100) plane and another part was attached to a Focused Ion Beam lift-out Cu-grid and Ar ion-milled with a Fishione Model 1010 to perform observations along the [001] direction of $\beta$-Ga$_2$O$_3$. Plasma cleaning was applied on the samples directly before the TEM investigations with a Fishione Model 1020.

Classical Molecular Dynamics (MD) simulations were performed using the Large-scale Atomic/Molecular Massively Parallel Simulator (LAMMPS) code [45], employing a newly-developed machine-learnt interatomic potential of the Ga$_2$O$_3$ system [46] based on a tabulated Gaussian approximation potential (tabGAP) [47], which has been previously shown to accurately describe several properties of the $\alpha/\beta/\gamma/\delta/\kappa$ polymorphs [46,48]. The simulation cell consisted of a 320000-atoms $\beta$-Ga$_2$O$_3$ cell (20 × 40 × 20 conventional cells, corresponding to the cell dimensions of about 242 Å × 123 Å × 118 Å). Periodic boundary conditions were employed along the [010] and [001] direction, while the direction perpendicular to the (100) plane was subjected to non-periodic boundary conditions. We defined three layers parallel to the surface in the cell, following the work of J. Wu et al. [49] (see the Supporting Information for a schematic). A 10-Å boundary layer of fixed atoms was used in order to fix the position of the cell, followed by a 20-Å thermostat layer — NVT (number, volume, temperature), i.e, the canonical ensemble —which was modelled with a Nosé-Hoover thermostat at 300 K [50]. The goal of this layer is to simulate the experimental conditions where the energy introduced by the impinging ions will be dissipated by the pristine regions of the sample or the substrate. Finally, the remaining ~220-Å layer corresponded to atoms that interact directly with the implanted ions and followed the microcanonical ensemble — NVE (number, volume, energy). The projectiles consisted of Ga atoms that were initialized at random positions in the plane 20 Å above the surface (in order to avoid direct interaction with the surface atoms), with a kinetic energy of 10 keV. This choice is convenient, as the MD potential already addresses the interaction between the impinging Ga atoms and the Ga and O atoms of the lattice, and produces a similar number of vacancies per unit length as those obtained experimentally with a 250 keV Cr ion beam, ensuring the validity of the simulation approach, although the range is much smaller (see the Supporting Information for more details). The channeling effect was avoided by setting an angle of 7° with respect to the direction perpendicular to the (100) plane. The simulation was then run for 20 ps using an adaptive time step [51], which is dynamically adjusted to ensure that the energy and position of the atoms do not vary more than a given amount, based on the current values for the velocities and forces. Afterwards, the simulation is run for an additional 20 ps using the NVT ensemble at 300 K on the implantation region, with a time step of 0.001 ps. The simulation time was chosen so as to allow each collision cascade to fully develop and relax before the next impact. The electronic stopping of the ions was modelled as a friction term applied to the atoms with kinetic energies above 10 eV, in order to obtain an accurate estimation of the penetration depth [48]. The structure visualization was done using the Open Visualization Tool (OVITO) [52]. Complementary overlapping cascade simulations were performed with a different simulation scheme employing periodic boundary conditions. These simulations used $\beta$-Ga$_2$O$_3$ cells containing 81920 atoms within a simulation box of approximately 97 Å × 99 Å × 94 Å. In each iteration, a Ga or O atom is randomly selected as the primary knock-on atom (PKA). To ensure consistency, the entire cell is translated and wrapped at periodic boundaries, positioning the PKA at the center of the cell. The PKA is assigned a kinetic energy of 1.5 keV, with a uniformly random momentum direction. The collision cascade process is simulated within a microcanonical ensemble (NVE) for 5000 MD steps. As before, an adaptive time-step algorithm was employed to maintain numerical stability and electron-stopping frictional forces are applied to atoms with kinetic energies above 10 eV. Following the initial 5000 MD steps, the simulations proceed in a quasi-canonical ensemble. A Langevin thermostat is applied to atoms within 7.5 Å of the simulation box boundaries (which are redefined at each iteration) for 10 ps at 300 K. To release the stress in the defected cell, an additional relaxation step is carried out in an isothermal-anisobaric (NPT) ensemble at 0 bar and 300 K for 2 ps.

In order to directly compare the MD simulations with the experimental results, RSM maps were simulated employing the kinematic approximation of X-ray diffraction to the final configuration of the cell [53,54]. The intensity corresponding to the reciprocal space vector $\boldsymbol{Q}$ is given by:

$$I(\boldsymbol{Q}) \propto \left|\sum_{n} f_n(|\boldsymbol{Q}|) \exp(i\boldsymbol{Q} \cdot \boldsymbol{r}_n)\right|^2, \qquad (1)$$

where $\boldsymbol{r}_n$ is the position of the $n^{\text{th}}$ atom in the simulation cell and $f_n(\boldsymbol{Q})$ is the atomic scattering factor, which was modelled in this work using the Cromer-Mann parametrization with the tabulated coefficients for $Ga^{3+}$ [55] and $O^{2-}$ [56].

## 3. Structural data

### 3.1. Ion-beam-induced exfoliation

A (100)-oriented $\beta$-Ga$_2$O$_3$ sample was implanted with a fluence of $1.0\times10^{15}$ cm$^{-2}$ using 250 keV Cr$^{2+}$ ions, with a flux lower than ~$1.0\times10^{12}$ cm$^{-2}$ s$^{-1}$. Surprisingly, after the implantation, it was observed that a thin surface layer had been exfoliated and self-rolled along the $b$-axis ([010] direction), forming several microtubes. A Scanning Electron Microscopy (SEM) image of one of these tubes is shown in Fig. 1 (a). The self-rolling phenomenon is likely a result of the particular structure of the material, as both the (100) and (001) planes are easy-cleavage planes and the [010] direction corresponds to a well-reported crystalline habit line of $\beta$-Ga$_2$O$_3$ [57].

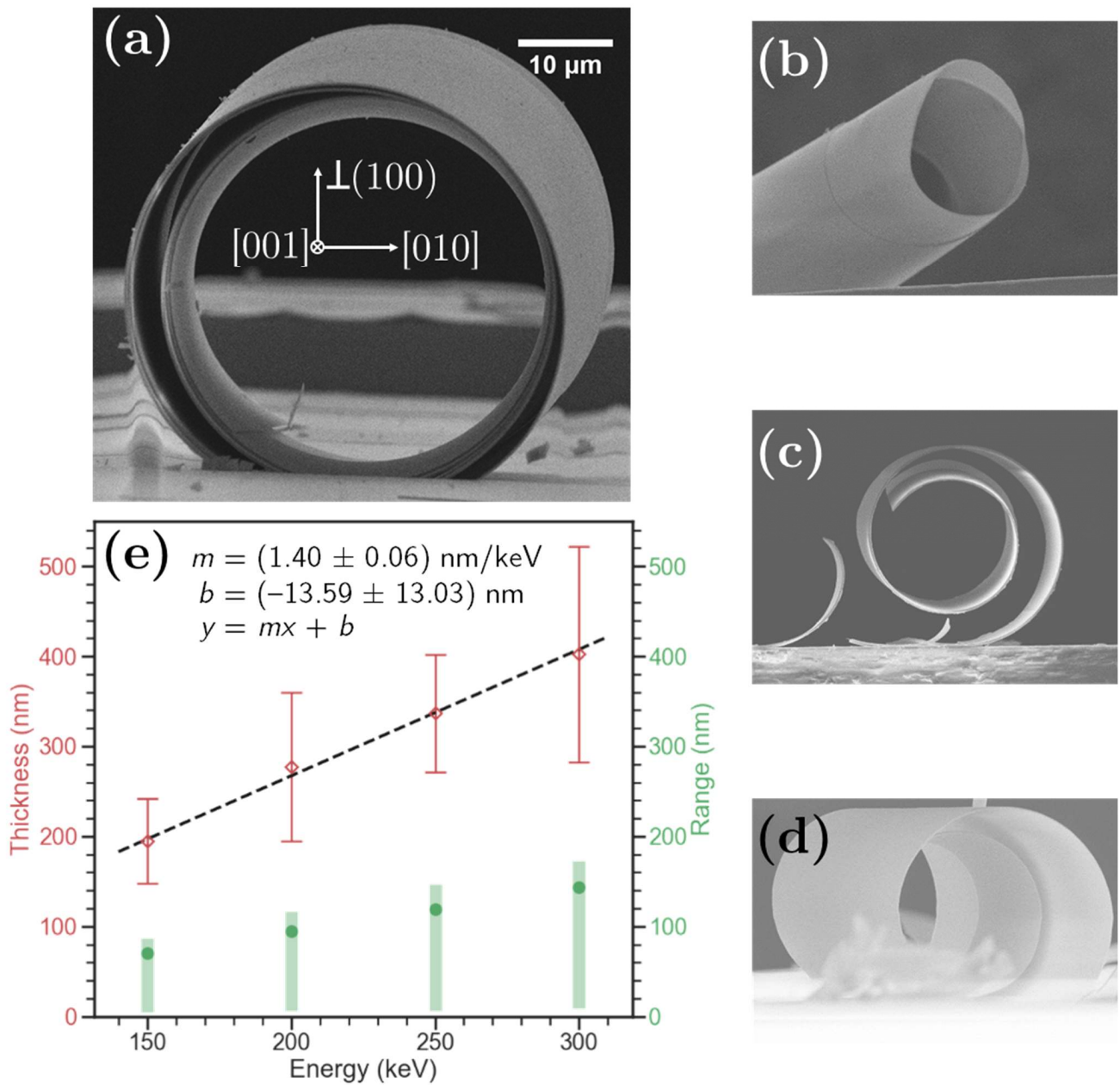

**Fig. 1** | (a) SEM image of a typical microtube produced by $Cr^{2+}$ implantation. The SEM images in panels (b), (c) and (d) show microtubes produced with $Co^{2+}$, $Cu^{2+}$ and $Al^{2+}$ ions, respectively. (e) Microtube thickness as a function of the energy of the implanted $Cr^{2+}$ ions (in red squares), with green bars denoting the standard deviation, along with their expected range according to SRIM Monte Carlo simulations (in green circles), with the error bars denoting the full width at half maximum of the vacancy distribution. The dashed black line in panel (e) corresponds to the least-square linear fit of the experimental data.

After their formation, the microtubes were transferred to a (100)-oriented Si substrate and annealed in air at 500 °C, on a conventional hotplate. Remarkably, the microtubes spontaneously unrolled during annealing, yielding flat nanomembranes that adhered to the Si substrate. A video of the unrolling phenomenon is available as Supporting Information. This process for producing microtubes and nanomembranes has been patented [27].

The key parameters that can be controlled during the implantation are the projectile species, its energy, fluence and flux (i.e., the beam current), as well as the implantation temperature, making it critical to assess their influence on the self-rolling phenomenon. In particular, it was possible to successfully reproduce the microtube formation process provided that the implantation flux was kept $<1.0\times10^{12}$ cm$^{-2}$ s$^{-1}$. This observation may be related either to a flux-dependent defect formation dynamics, which has been reported for $\beta$-Ga$_2$O$_3$ [19,58,59], or with the local heating of the sample during the implantation, which could be particularly relevant for $\beta$-Ga$_2$O$_3$, due to its low thermal conductivity [8,60].

In order to confirm the reproducibility and, at the same time, assess the impact of the fluence, a set of samples were implanted using 250 keV Cr$^+$ ions, with a constant flux of $2.0\times10^{10}$ cm$^{-2}$ s$^{-1}$ with fluences of $6.0\times10^{12}$, $1.5\times10^{13}$ cm$^{-2}$, $4.0\times10^{13}$ cm$^{-2}$, $1.0\times10^{14}$ cm$^{-2}$, $2.0\times10^{14}$ cm$^{-2}$ and $6.0\times10^{14}$ cm$^{-2}$ at RT. It was observed that microtubes formed only for fluences $\geq1.0\times10^{14}$ cm$^{-2}$ (optical microscopy images available in the Supporting Information), corresponding to ~0.3 displacements per atoms (dpa) at the maximum of the vacancy profile obtained from SRIM (more details in the Supporting Information). Thus, this fluence (and corresponding dpa) may be considered as an approximate exfoliation threshold, keeping all other parameters constant.

After establishing the limits in the implantation flux and fluence, the question of the role of the ions was addressed. The ion energies and fluences were tuned to yield defect profiles similar to the those obtained by implanting 250 keV Cr$^{2+}$ ions, according to SRIM. Using this approach, it was possible to successfully produce microtubes with various ions, including Co, Cu and Al, as shown in Figs. 1 (b), (c) and (d), as well as Fe and W (not shown). The possibility to implant different ions greatly enhances the potential of this process, since the optical, structural, mechanical, electrical or magnetic properties of the membranes can be tailored to fit the envisaged application, considering a wide range of dopants. Moreover, it is also a clear indication that the process is triggered by the implantation-induced defects, rather than chemical effects depending on the implanted ion.

In order to address the effect of the implantation energy on the produced microtubes, multiple implantations of Cr$^+$ ions were performed under similar conditions, but with varying energies. Using higher implantation energies increases the ion projected range, at the cost of greater straggling. Fig. 1 (e) shows the microtube wall thickness (i.e., the thickness of the rolled layer) and projected ion ranges as a function of the implantation energy, revealing a linear trend in the estimated energy range. The wall thicknesses of the produced tubes were measured by SEM imaging performed in cross-sectional view. From the experimental point of view, this measurement is quite challenging, due to the ever-present misalignment between the view direction and axes of the tubes. As such, there is a risk of overestimating (but never underestimating) the thicknesses in Fig. 1 (e). Therefore, these results are interpreted as an upper bound for the tube wall thickness. Importantly, for each point in Fig. 1 (e), ~30 microtubes were measured, to obtain statistically relevant sets of data, from which the average values and standard deviations were calculated.

The results in Fig. 1 (e) show that thicker tube walls can be obtained with increasing energies, in agreement with expectations, since the implantation energy defines the ion penetration depth. Importantly, these data confirm an opportunity to obtain microtubes with controlled thicknesses, within an uncertainty range. Notably, while both the thickness and the ion projected range trends scale similarly with the energy, the former is larger than the latter, even considering the contribution of straggling. As such, the implanted region always remains inside the tube walls, making the doping feasible, both in the case of the microtubes shown in Fig. 1 and in the case of the subsequently unrolled membranes.

### 3.2. Disorder-strain interplay

In order to assess the stain profiles induced by the ion implantation, the samples below the exfoliation fluence threshold were measured by High-resolution X-ray Diffraction (HRXRD). Fig. 2 (a) shows the $2\theta - \omega$ scans about the 400 symmetric reflection.

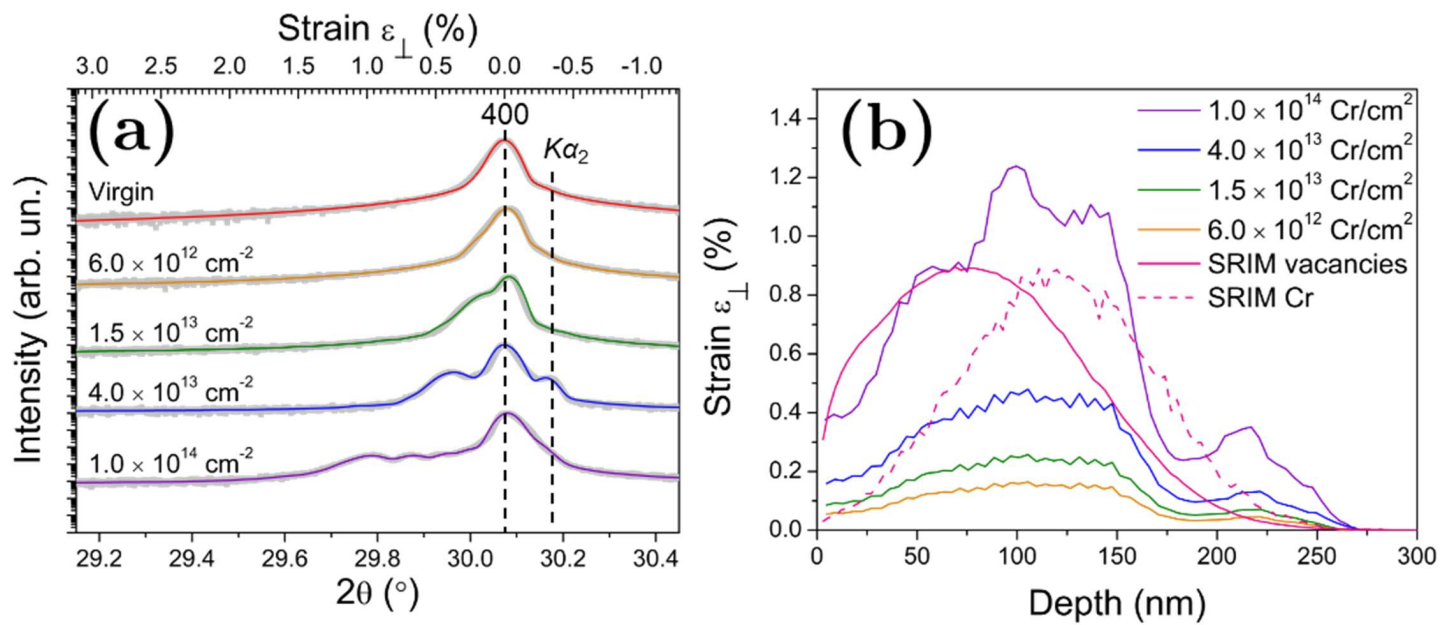

**Fig. 2** | (a) $2\theta - \omega$ scans about the 400 symmetric reflection (symbols), together with the corresponding MROX fits (lines), for different implantation fluences of 250 keV, in comparison with the virgin sample. The curves were shifted vertically for visual clarity. The minor peak on the right-hand side of the main peak is due to the presence of the partially unfiltered Cu K$\alpha_2$ X-ray line. The upper scale represents the angle-to-strain conversion. (b) Strain profiles obtained from the data in panel (a), together with the Cr and vacancy distributions obtained with SRIM (plotted in arbitrary units).

The HRXRD results reveal a broad shoulder for implanted samples at lower $2\theta$ values with respect to the diffraction peak of the virgin sample. This corresponds to an increase of the (400) interplanar distance $d = \frac{a \sin \beta}{4}$, where $a$ is the first lattice constant and $\beta$ is the monoclinic angle, allowing the perpendicular strain $\varepsilon_\perp = \frac{d-d_0}{d_0} =$ $= \frac{\sin\theta_0 - \sin\theta}{\sin\theta}$ to be calculated, where $\theta_0$ and $d_0$ are the Bragg angle and interplanar distance of the virgin region of the sample, respectively. As such, we observe an out-of-plane expansion (positive strain), increasing with the implantation fluence. Note that, since the interplanar distance depends not only on $a$, but also on the monoclinic angle $\beta$, this strain does not translate entirely to the strain $\varepsilon_a$ along the $a$ lattice parameter (i.e., $\varepsilon_a \neq \varepsilon_\perp$).

The perpendicular strain profiles were extracted from the aforementioned fits to the experimental data and are plotted in Fig. 2 (b), along with the Cr and vacancy distributions (in arbitrary units) obtained from SRIM. The shapes of the strain profiles and the SRIM vacancy distribution are in good agreement. Specifically, the strain distribution achieves a maximum at ~100 nm and extends up to ~250 nm, which is slightly deeper than the simulated vacancy profile. The maximum strain increases with the fluence, while the shape of the strain profile remains very similar.

While symmetric $2\theta - \omega$ scans probe the strain along the direction perpendicular to the (100) surface plane, the strain in the in-plane directions was assessed using RSM about the $710$ and $80\bar{1}$ reflections, in the grazing exit geometry, supplying information about strain along the [010] and [001] directions, respectively. These RSM are shown in Figs. 3 (a) and (b), for the sample implanted with a Cr fluence of $4.0\times10^{13}$ cm$^{-2}$ (~0.1 dpa). The algebraic expressions for the reciprocal space quantities were derived using a Mathematica script [61].

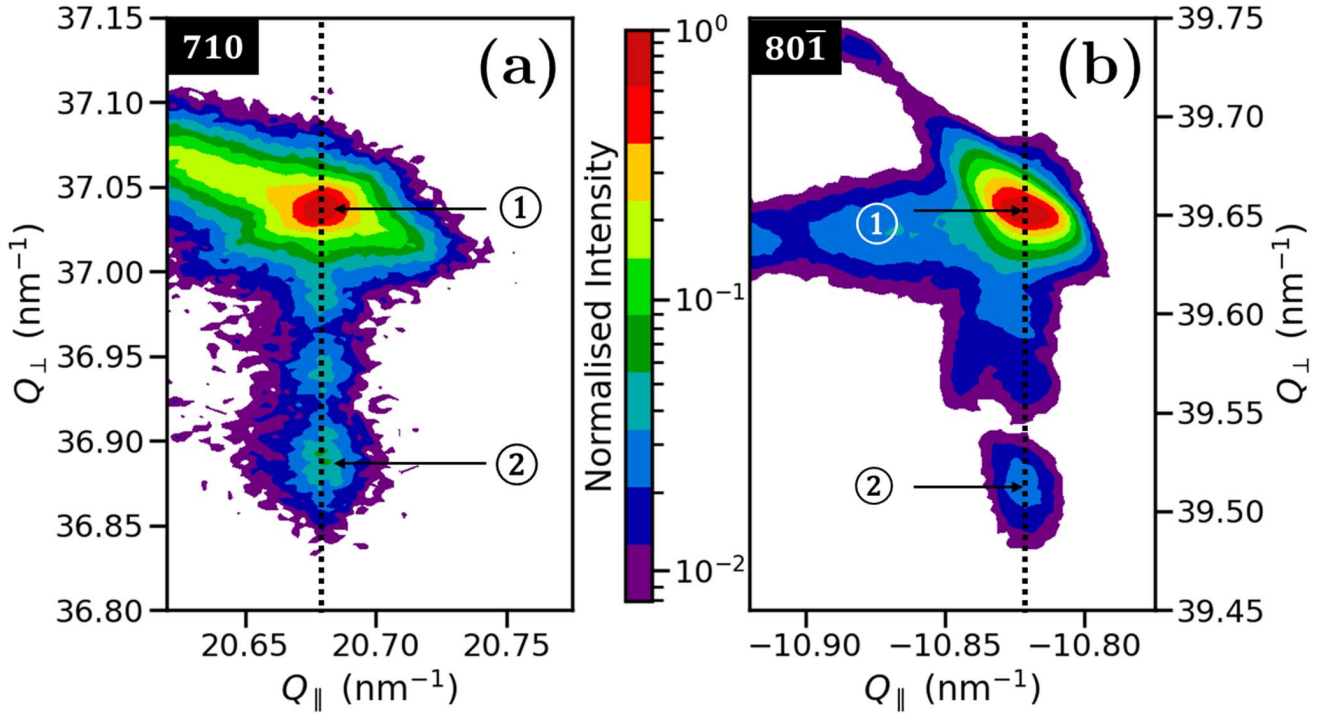

**Fig. 3** | Experimental reciprocal space maps obtained about the (a) $710$ and (b) $80\bar{1}$ reflections of the sample implanted with a fluence of $4.0\times10^{13}$ cm$^{-2}$ of 250 keV Cr. The dashed lines indicate the position corresponding to the parallel components of the scattering vector. The arrows marked with ① indicate the maxima of the peaks associated with the pristine region of the sample, whereas those marked with ② indicate the maxima of the peaks corresponding to the greatest perpendicular strain.

The shapes of the reciprocal space maps are unusual and broad, which we attribute to the uneven (stair-like) surface of the sample, occurring due to the easy cleavage (100) plane. However, it can be clearly seen from Fig. 3 that the position of the parallel component of the scattering vector does not change, which suggests that both the $b$ and $c$ lattice parameters remain the same after the ion implantation, i.e., there is no induced strain along these directions. On the other hand, the decrease of the perpendicular component of the scattering vector suggests an increase of the interplanar distance associated with the {100} family of planes, in agreement with the $2\theta - \omega$ scans (see the Supporting Information for a more detailed characterization of the lattice parameters based on the RSM). Such a strain state is in agreement with what is typically expected of implanted single-crystals, as the in-plane lattice parameters of the damaged layer are fixed by the presence of the undamaged regions of the sample, introducing stresses in the lattice. Due to the Poisson effect, these stresses contribute to the enhancement the out-of-plane expansion, as the surface of the sample is free to change [62].

The evolution of the defect profiles as a function of the fluence was also investigated by RBS/C. Fig. 4 (a) shows the RBS spectra of a randomly-oriented sample, as well as the spectra pertaining to the pristine and as-implanted samples with the beam oriented along the [201] axis, which is nearly perpendicular to the (100) plane (~0.1°).

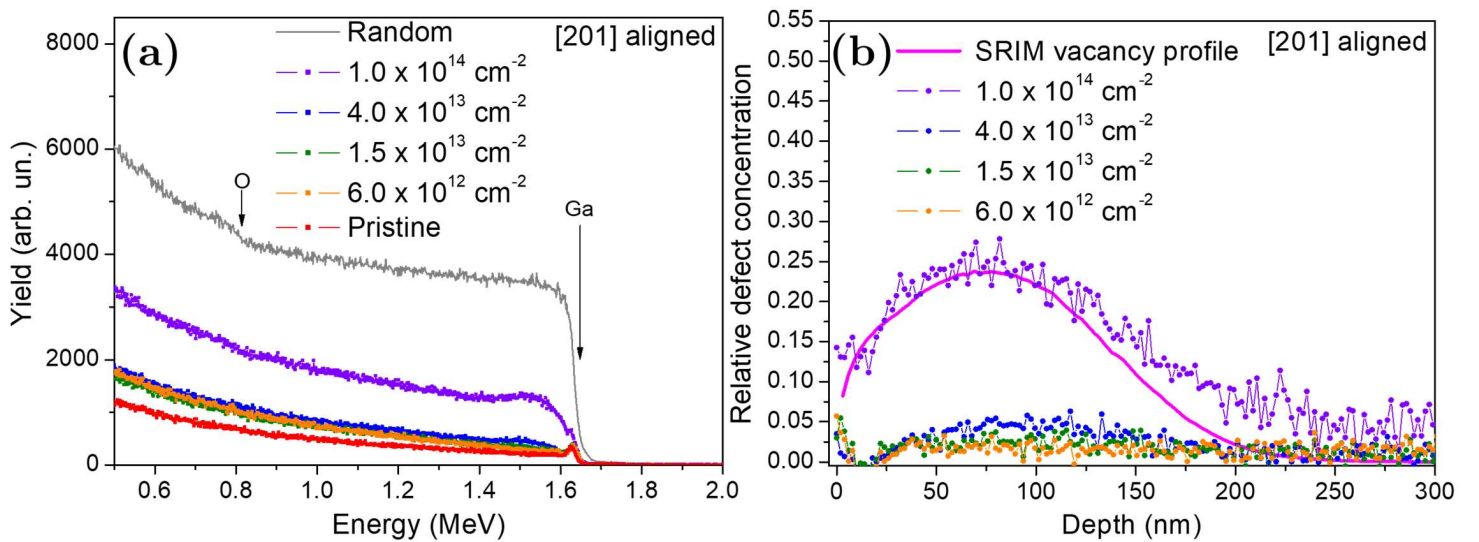

**Fig. 4** | (a) RBS/C spectra of the pristine and implanted samples, with the beam oriented along the [201] direction, as well a random spectrum. (b) Relative defect concentration profiles extracted according to the two-beam approximation, as well as the vacancy profile calculated from SRIM.

All the channeled spectra in Fig. 4 (a) show a clear peak due to direct backscattering at the surface. For the pristine sample, the yield of this peak is ~5% of that of the random spectrum, which is slightly higher than what is found in Si or GaN samples [36], and suggests the presence of some defects already in the pristine sample. After implantation, apart from this surface peak, it is possible to identify a clear structure for energies between 1.5 and 1.6 MeV, due to the increase of the lattice disorder, corresponding to direct backscattering from point defects, such as interstitials, within the channel. Moreover, an enhancement of the backscattering yield for low energies is also visible, and corresponds to the dechanneling phenomenon that will ultimately lead to backscattering in deeper regions of the sample, thus contributing to the low-energy portion of the spectrum.

The defect profiles extracted from the RBS/C spectra are shown in Fig. 4 (b). It is clear that they are in agreement with the vacancy profiles obtained from SRIM. Importantly, even though the defect concentration increases as a function of fluence, it remains well below a fully disordered or amorphized state. This is also in line with previous studies for (100)-oriented crystals implanted with Eu, where K. Lorenz et al. report the amorphization of the sample at the surface for fluences above $1.0\times10^{15}$ cm$^{-2}$ (~11 dpa) [41]. On the other hand, M. Sarwar et al. observed amorphization of (010)-oriented $\beta$-Ga$_2$O$_3$ crystals implanted with Sm fluences above $3.0\times10^{15}$ cm$^{-2}$ (~20 dpa) [63], while E. Wendler et al. did not observe amorphization in samples with the same orientation and implanted with P, Ar and Sn ions with fluences up to $2.0\times10^{15}$ cm$^{-2}$ (up to ~3.2 dpa) [39]. The former group also reported surface amorphization in Yb-implanted ($\bar{2}$01)-oriented crystals [64]. Moreover, the defect concentration levels used in this study also remain well below the potential disorder-induced ordering polymorph transition reported recently in the Ga$_2$O$_3$ system at ~0.5 dpa [48,65,66].

Finally, in order to investigate the microtube structure on the nanoscale (S)TEM imaging and EDX spectroscopy were performed. Fig. 5 shows the data for a microtube obtained with a 250 keV Cr$^+$ implantation with a fluence of $1.0\times10^{14}$ cm$^{-2}$. These observations were conducted along the [001] $\beta$-Ga$_2$O$_3$ crystallographic direction. The average thickness of the successive rolled-up layers ranged from 400 up to 500 nm [see Figs. 5 (a) and (b)], in good agreement with Fig. 1. Selected-area electron diffraction (SAED) patterns [Fig. 5 (c)], along with high--resolution STEM and TEM images [Figs. 5 (d)–(f))], indicated that the microtubes remained as $\beta$-Ga$_2$O$_3$ single crystals. In particular, the results show no signs of the disorder-induced phase transformations due to the relatively low fluences, which remain below the dpa threshold level for such transition [67]. Meanwhile, it seems that a high-density of irradiation-induced defects are accumulated at the (100) cleavage planes, resulting in the formation of short- and long-lateral length stacking faults that possibly promote the detachment and rolling of the surface layers along the [010] direction. No chemical inhomogeneities or signs of amorphization were observed either.

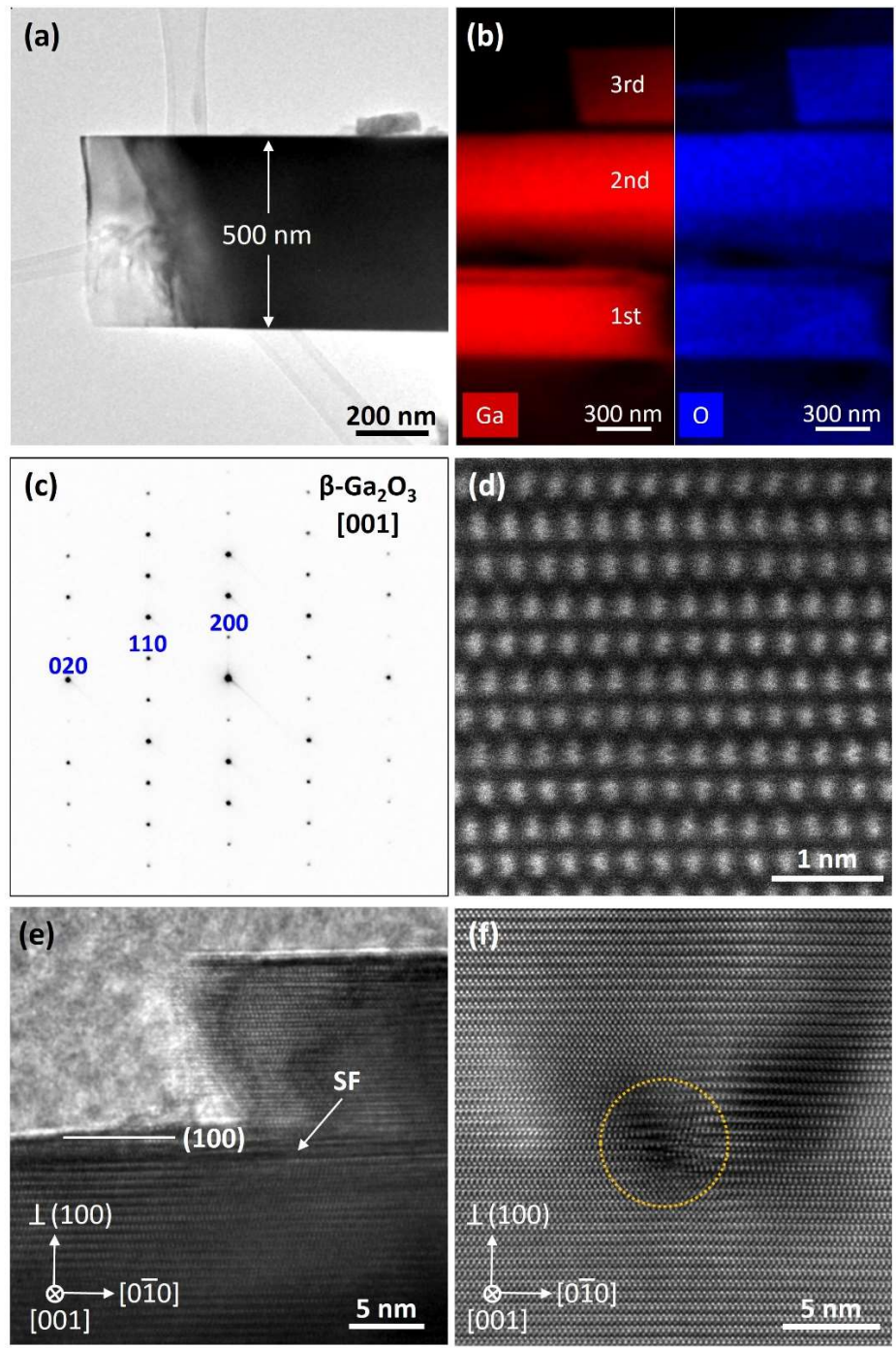

**Fig. 5** | (S)TEM nanoscale investigations of a $\beta$-Ga$_2$O$_3$ microtube along the [001] cross-sectional direction (obtained with a 250 keV Cr$^+$ implantation with a fluence of $1.0\times10^{14}$ cm$^{-2}$). (a) Low-magnification TEM image illustrating part of the detached microtube and (b) EDX maps of parts of three successive rolled-up layers of the microtube. (c) SAED pattern and (d) high-resolution HAADF-STEM image from the implanted area of the microtube, showing no polymorph transitions or amorphization. (e) and (f) high-resolution images illustrating the defect accumulation along the (100) cleavage planes, revealing extended stacking faults (SF) along the (100) planes that possibly contribute to the exfoliation and rolling of the surface layers along the [010] direction.

## 3.3. Disorder and strain evolution with annealing

As mentioned before, annealing the microtubes promotes the relaxation of the strain and leads to the unrolling of the tubes, yielding a flat nanomembrane. In order to systematically study this process, we monitored the evolution of the strain and defect profiles for a sample implanted with a fluence just below the exfoliation threshold discussed in section 3.1, namely, a sample implanted with 250 keV Cr, with a fluence of $5.0\times10^{13}$ cm$^{-2}$. The rapid thermal annealing steps were performed in an N$_2$ atmosphere.

Fig. 6 (a) shows the $2\theta - \omega$ scans about the 400 symmetric reflection and their respective fits, after annealing at temperatures between 250 and 1000 °C.

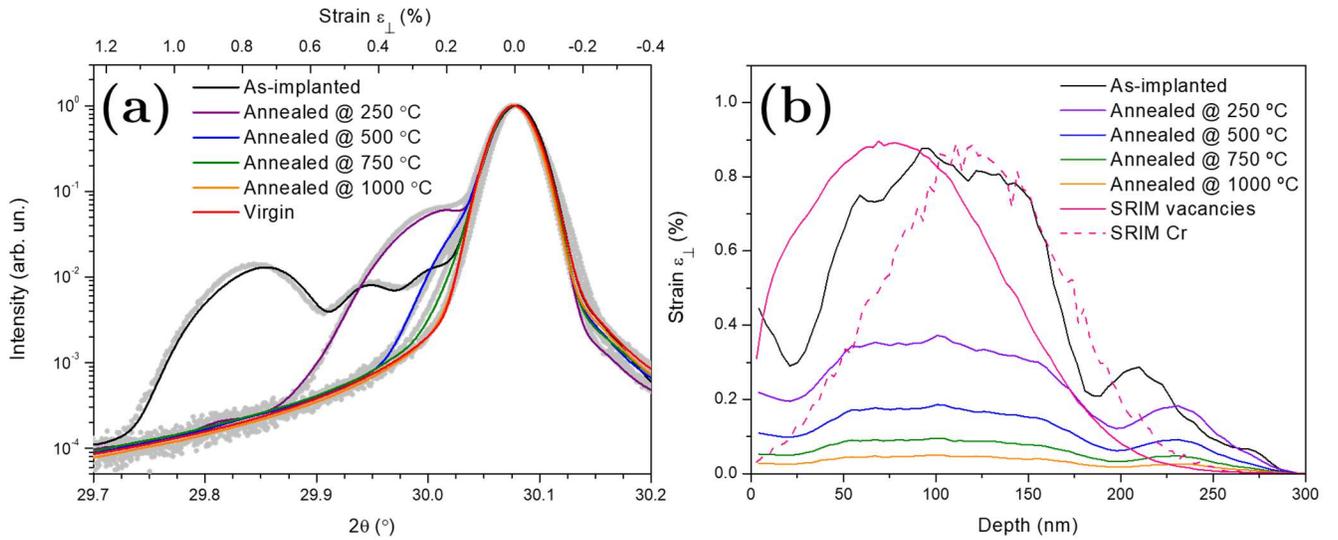

**Fig. 6** | (a) $2\theta - \omega$ scans about the 400 symmetric reflection (grey symbols), and the corresponding fits (colored lines), as a function of the annealing temperature, for the sample implanted with a fluence of $5.0\times10^{13}$ cm$^{-2}$ of 250 keV Cr ions. The upper scale represents the angle-to-strain conversion. (b) Strain profiles obtained from the fits in panel (a), together with the Cr and vacancy distributions obtained from SRIM (plotted in arbitrary units).

The data for the as-implanted sample in Fig. 6 are very similar in shape to the $4.0\times10^{13}$ cm$^{-2}$ curve shown in Fig. 2, thus confirming that the fluence is just below the exfoliation threshold, as desired for this study. Importantly, upon annealing, the structure on the low-angle shoulder of the main peak is gradually reduced, which is related with the strain relaxation in the lattice. Indeed, a reduction of the maximum strain by a factor ~2 is observed already upon annealing at 250 °C. Increasing the annealing temperature leads to a further strain relaxation. Remarkably, for an annealing temperature of 1000 °C the recovery is nearly complete. The strain profiles obtained from the fit, shown in Fig. 6 (b), are both qualitatively and quantitatively similar to the ones in Fig. 2. Such remarkable lattice recovery is also in agreement with data collected by Raman spectroscopy from annealed nanomembranes (see the Supporting Information), thus confirming their high crystallinity, which is akin to bulk single-crystals.

Finally, Fig. 7 shows the RBS/C data for the as-implanted and the one annealed at 250 °C.

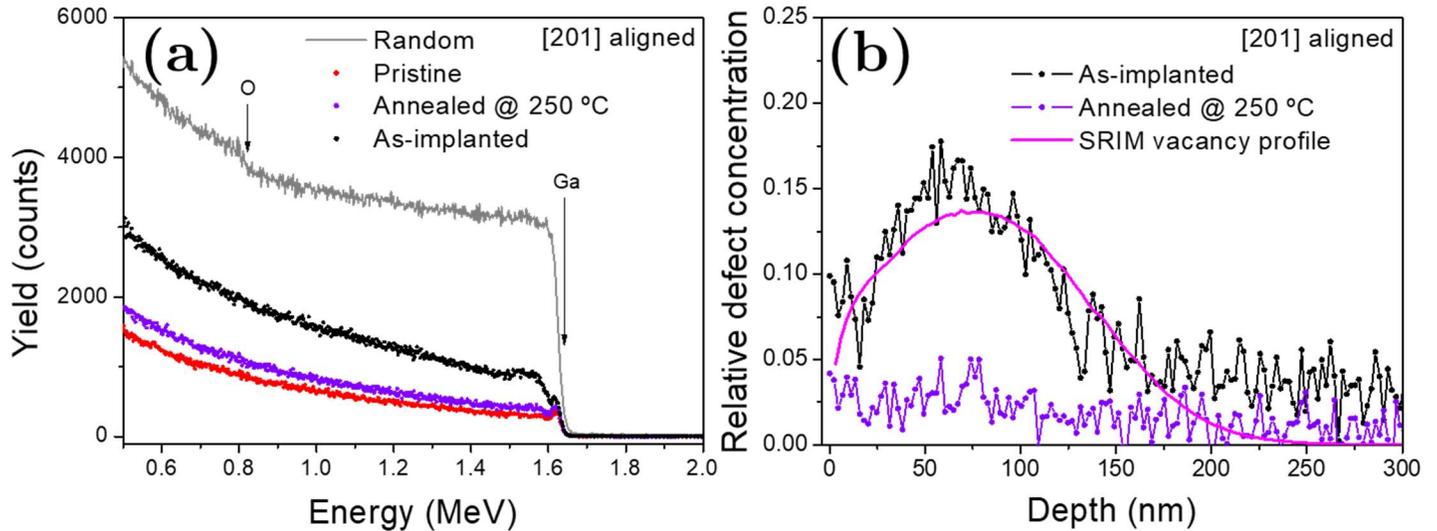

**Fig. 7** | (a) RBS/C spectra of the pristine, as-implanted ($5.0\times10^{13}$ cm$^{-2}$) and annealed at 250 °C samples, aligned with the [201] direction, in comparison with a random spectrum. (b) Relative defect concentration profiles extracted according to the two-beam approximation, as well as the vacancy profile calculated from SRIM.

Akin to the data in Fig. 4, the defect profile of the as-implanted sample is observed to closely follow the vacancy profile obtained from SRIM. Moreover, upon an annealing at a temperature of 250 °C, the defect profile essentially vanishes, indicating a defect concentration below the RBS/C detection limit of the technique, and confirming the remarkable recovery capability of the lattice already suggested by XRD.

## 4. Atomistic simulations and discussion

In order to inspect the atomic origins of strain and stress, classical MD simulations were performed using the LAMMPS software, as detailed in section 2, using dpa values comparable to the one used experimentally. Specifically, we used 20 sequential collision cascades induced by 10 keV Ga ions, yielding a disorder level on the simulation cell that is similar to that of the sample implanted with a fluence of $1.5\times10^{13}$ cm$^{-2}$ 250 keV Cr ions obtained experimentally, i.e., ~3.5 vacancies/Å for each incident ion at the maximum of the vacancy profile, according to SRIM (see the Supporting Information for more details), corresponding to ~0.05 dpa. Thus, Fig. 8 shows a snapshot of the simulation cell projected along two different directions after these 20 overlapping collisions, and how the defects compare with the SRIM vacancy profile (see the Supporting Information for an animated version of this visualization).

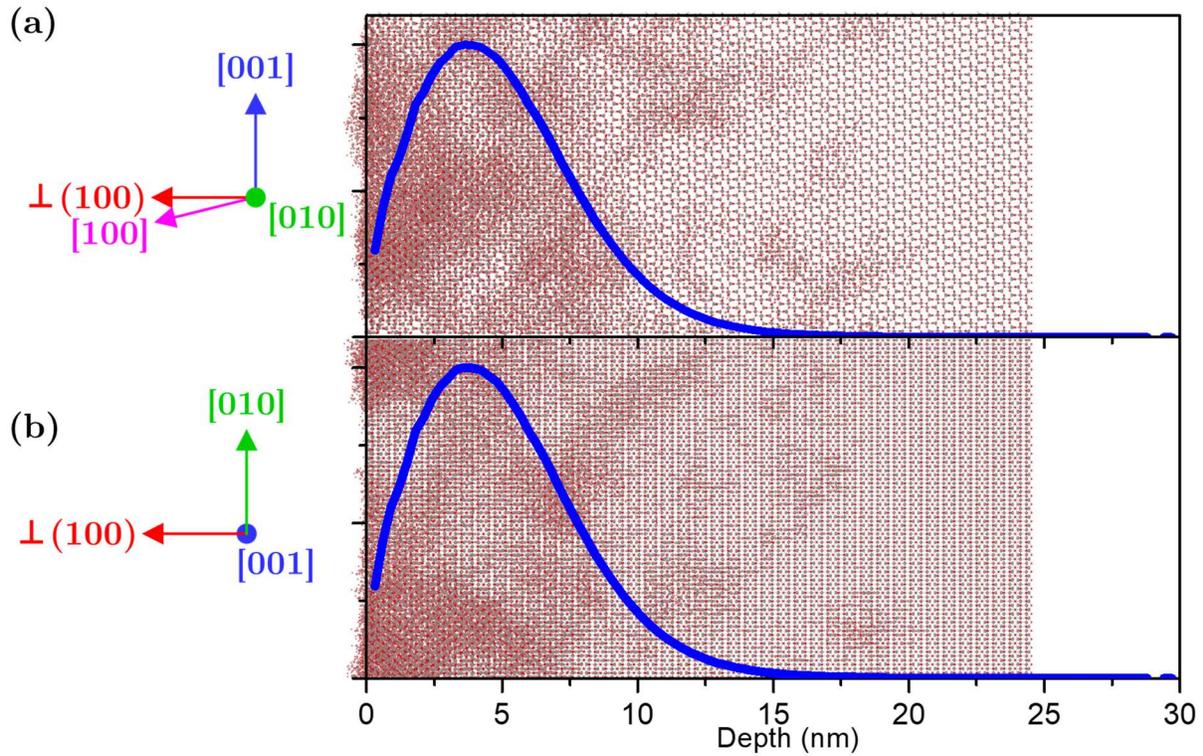

**Fig. 8** | Snapshots of the simulation cell after 20 overlapping 10 keV Ga ion impacts, as observed along the (a) [010] and (b) [001] directions. The blue curves correspond to the vacancy profiles obtained from SRIM.

Generally, Fig. 8 shows a good agreement between SRIM and the MD simulations. Indeed, most of the damage is produced in the region predicted by SRIM, although there are deeper defects in the MD simulations, attributed to the channeling phenomenon not accounted for by SRIM.

In order to compare with the experimental results, RSM were simulated on the vicinity of two different experimentally-accessible reflections $\overline{7}10$ and $80\overline{1}$ (see Fig. 3), which supply information on in- and out-of-plane components of strain, as shown in Fig. 9 (see the Supporting Information for more details). In agreement with the experimental RSM, it is clear for both maps that the parallel component of the scattering vector $Q_{\parallel}$ of the main peak is not altered by the implantation, suggesting that the in-plane lattice parameters $b$ and $c$ are not changed. On the other hand, the perpendicular component $Q_{\perp}$ is decreased after the irradiation, showing an increase of the out-of-plane interplanar distance. It should be noted that in experiment the implanted depth is small compared with the total volume probed by the X-rays. In other words, the experimental RSM contain information from both the damaged and pristine regions of the sample. As such, the RSM show a main peak associated with the virgin region and a structure for smaller values of $Q_{\perp}$ pertaining to the damaged region. For the simulated RSM, this is not the case, as the simulation cell is much smaller. In this case, the XRD signal comes almost entirely from the damaged region of the cell, leading to the shift of the whole peak when comparing the pristine and damaged systems. Moreover, many smaller interference fringes are observed, which arise due to the finite size of the cell.

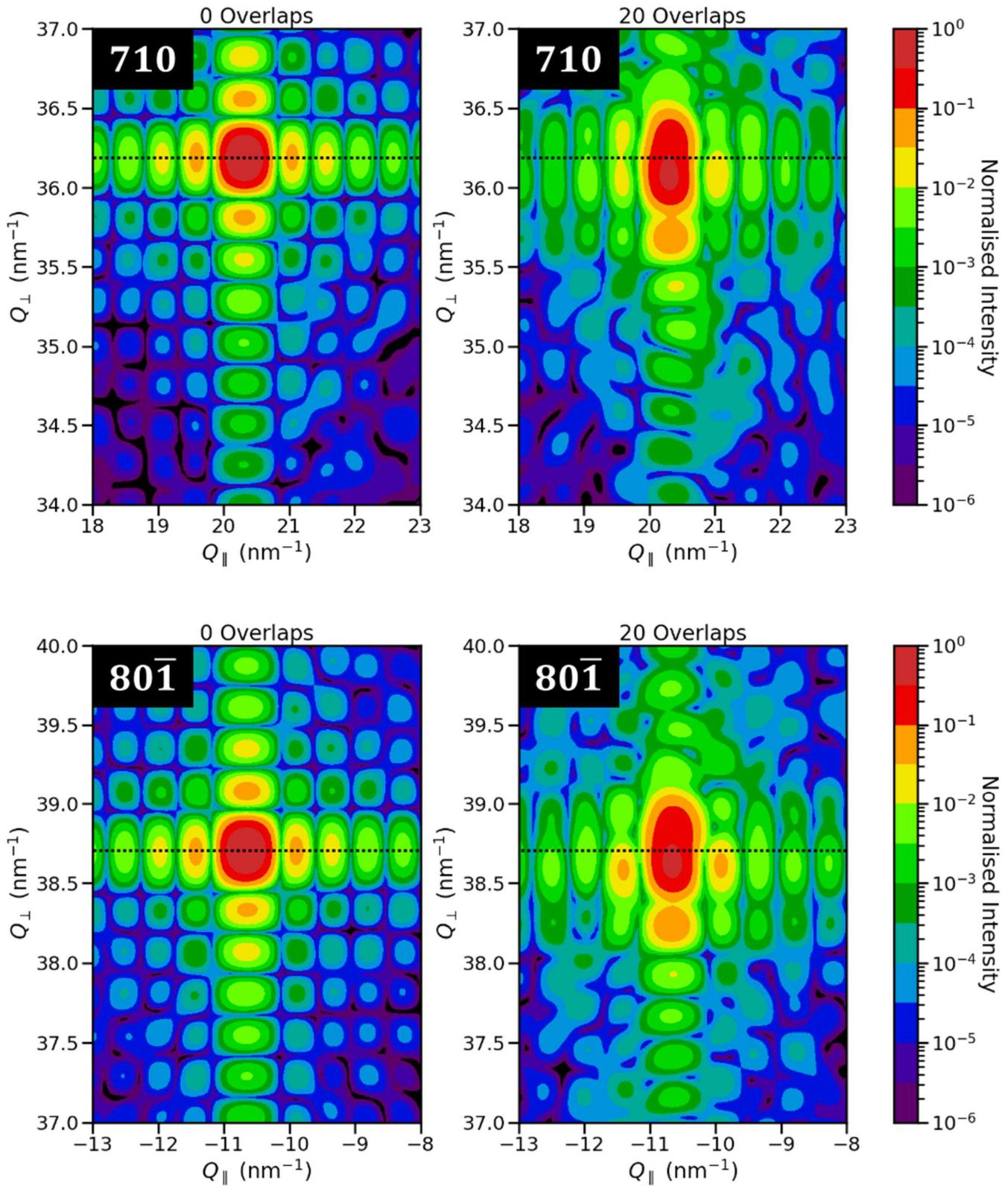

**Fig. 9** | Simulated reciprocal space maps obtained around the $710$ and $80\bar{1}$ reflections before and after 20 overlapping ion impacts. The dashed lines indicate the position corresponding to the perpendicular component of the scattering vector of the pristine cell.

It is also possible to simulate $2\theta - \omega$ scans by selecting an appropriate section of the RSM; for a symmetric reflection, $Q_\parallel = 0$ and $Q_\perp = \frac{4\pi}{\lambda}\sin\theta$ and $\omega = \frac{2\theta}{2}$ [68]. Using these relations, it is possible to obtain a simulated $2\theta - \omega$ scan about the 600 reflection, as shown in Fig. 10. The 600 reflection was chosen instead of the 400 reflection due to the enhanced angular resolution (i.e., the same strain value will correspond to a larger angular separation between the virgin and implantation peaks) for the out-of-plane strain.

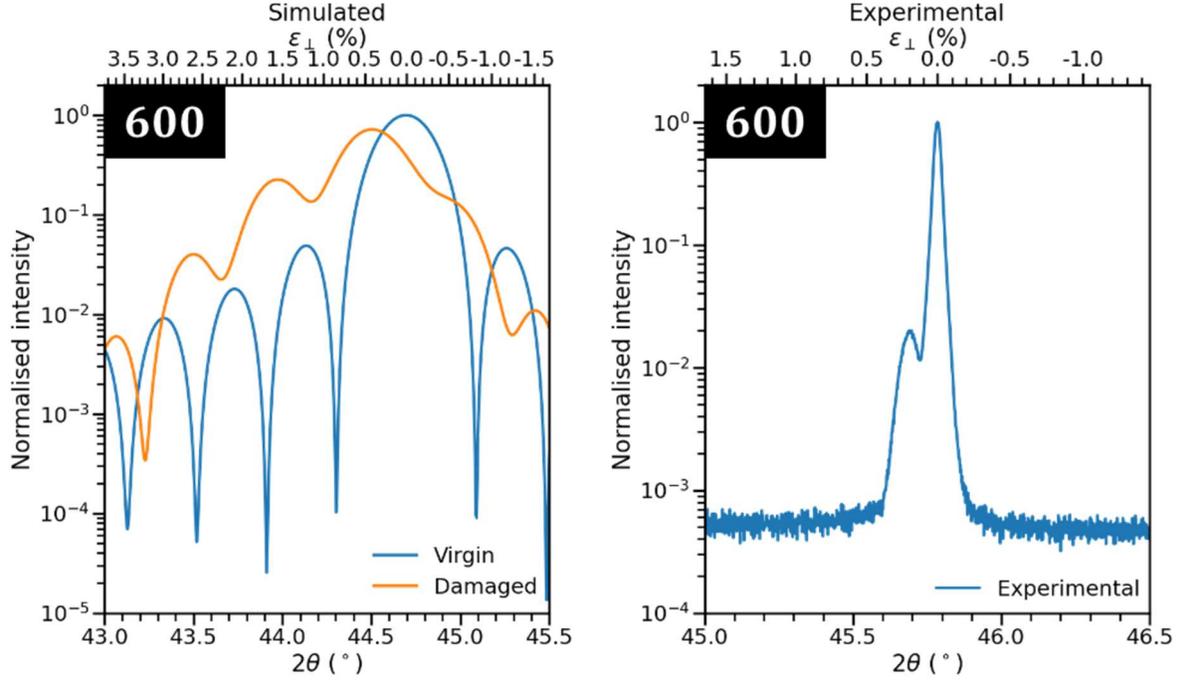

**Fig. 10** | Simulated (after 20 overlaps) and experimental (on the sample implanted with a fluence of $1.5\times10^{13}$ cm$^{-2}$) $2\theta - \omega$ scans about the 600 reflection. These fluence and overlaps correspond to ~0.05 dpa.

In spite of the visual difference between the simulated and experimental scans, as discussed above for the RSM, the results in Fig. 10 show an excellent agreement between the perpendicular strains calculated from the simulated and experimental $2\theta - \omega$ scans. Indeed, the position of the peak in the simulated diffractogram of the damaged sample is very close to the experimentally observed maximum strain of ~0.5%. This result also validates the employed methodology, where the experimental results are compared with simulations corresponding to similar dpa levels.

In order to better understand the microscopic strains and stresses induced by the implantation, the atomic finite strain tensor associated with each atom was calculated from Ovito, considering the deformation gradient tensor of the displacement between the reference and final configurations, according to [52,69]

$$\varepsilon_{ij} = \frac{1}{2}\left(\frac{\partial u_i}{\partial X_j} + \frac{\partial u_j}{\partial X_i} + \sum_k \frac{\partial u_k}{\partial X_j}\frac{\partial u_k}{\partial X_j}\right), \quad (2)$$

where $u_i$, $u_j$ and $u_k$ denote the $i$th, $j$th and $k$th components of the displacement vector and $X_i$ and $X_j$ denote the $i$th and $j$th components of the vector position in the coordinates of the reference (pristine) configuration. The extracted strain profiles, shown in Fig. 11 (a), correspond to the average of each strain component within slabs parallel to the surface, with a thickness of 2.85 Å. The virial stress tensor associated with each atom was calculated according to [49,70,71]

$$\sigma_{ij} = \frac{1}{V}\left[mv_i v_j + \frac{1}{2}\sum_{n} R_i(n)\, F_j(n)\right], \qquad (3)$$

where $m$ is the mass of the atom, $v_i$ and $v_j$ denote the $i$th and $j$th components of its velocity, $R_i(n)$ denotes the $i$th component of the (relative) position vector of the $n$th neighbour with respect to the considered atom and $F_j(n)$ denotes the $j$th component of the force exerted on the considered atom by its $n$th neighbour. The sum runs over the neighbours $n$ in the volume $V$, chosen as the volume enclosed by a sphere with a 5.0-Å radius (corresponding to the cut-off distance of the MD potential) surrounding that atom. The stress profiles, shown in Fig. 11 (b), correspond to the average of each stress component within slabs parallel to the surface, each with a thickness of 18 Å. Note that the strain and stress components in this work are expressed with respect to an orthonormal basis $\{e_1, e_2, e_3\}$, where $e_2$ and $e_3$ are unit vectors along the $b$- and $c$-axes, while $e_1$ is a unit vector parallel to the reciprocal lattice basis vector $a^*$, i.e. it is perpendicular to the (100) plane. Moreover, the convention used is such that a positive stress is tensile, while a negative stress is compressive. Table I summarizes the average values of the 6 components of the finite atomic strain and stress tensors along the corresponding profiles.

The simulated atomic finite strain profiles reveal that indeed the dominant component is the one perpendicular to the surface ($\varepsilon_{xx}$), while all the others are at least one order of magnitude smaller. This follows closely the previous discussion on the expansion along the out-of-plane direction. Regarding the stress profiles, it is possible to observe that the dominant components are $\sigma_{yy}$ and $\sigma_{zz}$ (along the [010] and [001] directions, respectively), with opposite signs.

The MD results are thus in excellent agreement with the discussion based on the experimental results of section 3, where we argued that the lack of strain along the in-plane directions can be explained by the presence of a latent stress. Interestingly, the sign of the two components is different, implying a tensile stress along [001] and a compressive stress along [010]. Under these circumstances, and considering the anisotropic nature of the monoclinic system, the results suggest that the thin implanted layer is subjected to an anisotropic in-plane stress state, while the stress along the direction perpendicular to the surface is zero, since the surface is free. In contrast, the in-plane directions along the $b$- and $c$- axes, however, are subjected to significant stresses with opposite signs. Conversely, the strain along these two in-plane directions is approximately zero, while the strain in the direction perpendicular to the surface is significant and positive. This means that the sample is free to expand in the direction perpendicular to the surface, while it is constrained to keep the $b$ and $c$ lattice parameters constants by the remaining pristine sample (i.e., the "substrate"), thus leading to an accumulation of stress in these directions.

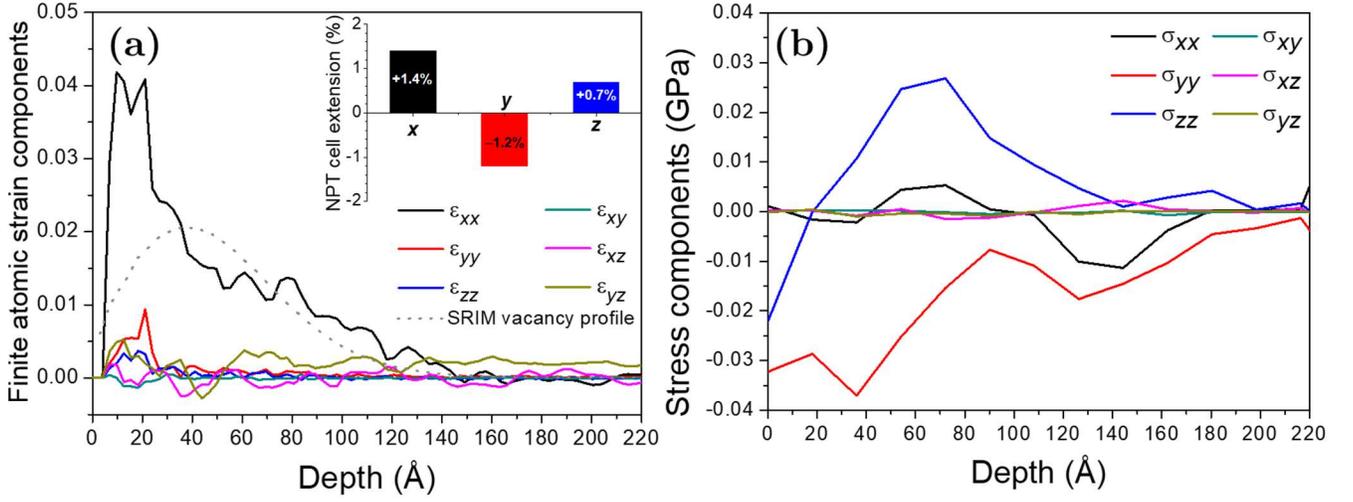

**Fig. 11** | Simulated atomic finite strain (a) and stress (b) components as a function of depth, after 20 overlapping impacts simulated in the NVT ensemble. The inset in (a) shows the cell extension along the $x$-, $y$- and $z$-directions, as simulated in the NPT ensemble.

| Components | Strain | Stress (GPa) |
|---|---|---|
| $\varepsilon_{xx}$ | $+6.88 \times 10^{-3}$ | $+3.831 \times 10^{-4}$ |
| $\varepsilon_{yy}$ | $+6.56 \times 10^{-4}$ | $-1.585 \times 10^{-2}$ |
| $\varepsilon_{zz}$ | $+2.73 \times 10^{-4}$ | $+5.211 \times 10^{-3}$ |
| $\varepsilon_{xy}$ | $-6.14 \times 10^{-5}$ | $-1.255 \times 10^{-4}$ |
| $\varepsilon_{xz}$ | $+1.61 \times 10^{-3}$ | $-1.378 \times 10^{-4}$ |
| $\varepsilon_{yz}$ | $-3.69 \times 10^{-5}$ | $-2.077 \times 10^{-4}$ |

**Table I** | Depth-averaged component-resolved atomic finite strain and stresses shown in Fig. 11.

To further investigate the strain state, additional anisotropic NPT ensemble MD simulations were employed, allowing to relax the stress induced by radiation defects in the cells, as described in section 2 above. According to the NPT-dpa model [72], 155 overlapping cascades were performed, corresponding to a maximum of ~0.05 dpa, in agreement with the calculations from SRIM Monte Carlo simulations for 10 keV Ga ions. This additional relaxation step in the NPT ensemble allows for the expansion of the cell to relax the stresses induced during the irradiation. The cell extensions along the $x$, $y$ and $z$ directions, shown in the inset of Fig. 11 (a), reveal a clear expansion along the $x$-direction and contraction along the $y$-direction, in agreement with the stresses calculated in the NVT ensemble and shown in Fig. 11 (b). Regarding the $z$-direction, a smaller expansion was observed, which suggests that the accumulated stress calculated in the NVT ensemble can, in a physical experiment, be locally relaxed at the expense of lattice deformation. This is particularly interesting, considering that $\gamma$-phase inclusions are often reported as a common defect in implanted $Ga_2O_3$ [73], and their formation is associated with a volume change [46]. Therefore, while the calculated $z$-stress in the NVT ensemble may be exaggerated, its positive sign agrees with the slight expansion observed during the anisotropic NPT relaxation step.

These observations can be considered in order to explain the observed rolling-up effect along the [010] direction. In fact, above a certain stress threshold, which in the current case occurs for fluences above $1.0 \times 10^{14}$ cm$^{-2}$, the surface layer detaches from the sample. The physical mechanism triggering the process at a certain depth is still unknown, but may be related to some pre-existent defect; likewise, the definition of the lateral dimensions of the tubes is still not fully understood. Once the layer becomes free, the stress along [010] compresses the implanted layer with respect to the deeper regions, resulting in a negative curvature radius and upwards bending. Moreover, this correlates well with the reported negative strain (contraction of the $b$ lattice parameter) that is observed experimentally in ion implantations performed perpendicularly to the (010) surface [58].

In the literature, it is possible to find different reports of ion-beam-induced bending or breaking effects, which have been studied in different materials. On the one hand, in the case of Si, Ge, GaAs and ZnO nanowires, it has been shown how the volume differences due to the mismatch in the vacancies and interstitials depth-profiles can lead to bending either in the direction of the ion beam or away from it [74–78]. In the case of Si membranes, M. Masteghin et al. showed how bi-material bending can be achieved in a controlled manner by ion implantation [79], proposing a model based on S. Timoshenko's paper on bi-metal thermostats [80]. Adopting this latter model, we consider the nanomembranes as bi-layers where the top layer (with thickness $t_f$) is the implanted region and the bottom layer (with thickness $t_s$) is pristine. From the experimental results shown above [Fig. 1 (a)], for an implantation with 250 keV Cr ions, the curvature radius is $|R| = 20$ μm, and the thickness of the wall is on average $t = 300$ nm. The implanted region corresponds to the first $t_f = 250$ nm, while the remaining unimplanted region ($t_s = 50$ nm) acts as the substrate, yielding a ratio $r_t = 5$ between the two. Considering that the elastic properties are not too affected by the implantation, the ratio between the Young's moduli for these two regions should be close to 1. Thus, Timoshenko's formula yields a strain of [79]

$$\varepsilon_b = -\frac{t}{6|R|} \frac{6 + 4(r_t + r_t^{-1}) + (r_t^{-2} + r_t^{-2})}{2 + r_t + r_t^{-1}} \approx -1.8\%, \tag{4}$$

which is of the same order of magnitude as the one obtained for the $y$-direction in the NPT MD simulations (shown in Fig. 11).

## 5. Conclusions

This work presents a novel method based on ion implantation that can be used to produce thin nanomembranes and microtubes of $\beta$-Ga$_2$O$_3$ with improved control compared to conventional mechanical exfoliation. The implantation-induced defects create a distribution of stresses and strains that ultimately lead to the rolling-up of the surface layer of a (100)-oriented single-crystal, thus creating a microtube. Remarkably, under annealing at moderate temperatures (~500 °C), these microtubes unroll themselves, which leads to the creation of a nanomembrane of high single-crystalline quality.

This method allows the thickness of the membranes to be controlled by varying the implantation energy, and has been successfully reproduced using different ion beams (in contrast to the SmartCut® process, that relies on the implantation of H or He to form gas bubbles), namely Al, Cr, Fe, Co, Cu and W. The thickness of the membranes is larger than the penetration range of the ions, so that this method can be seen as a 2-in-1 process, allowing doping and exfoliation in a single step. The physical mechanism governing the splitting and self-rolling phenomena are associated with a combination of the easy cleavage planes with the anisotropic nature of the monoclinic system. On the other hand, a key advantage of this process over conventional mechanical exfoliation is its scalability, which has the potential to facilitate its application in industry.

Specifically, this work includes a detailed experimental study combining XRD and RBS/C, which is used to characterize the strain and defect profiles induced by the ion implantation, as well as their relaxation/recovery upon annealing. MD simulations were performed to clarify the strain and stress accumulation, revealing an excellent agreement with the experimental results. The powerful combination of experiment and simulation allowed to advance a model to explain the ion-beam-induced exfoliation process by shedding light on the strain, stress and defect profiles induced by the implantation, as well as their role in the rolling-up phenomenon. Moreover, this work contributes to the establishment of $\beta$-$Ga_2O_3$ as a promising wide bandgap semiconductor, by proposing a new, scalable and industry-friendly process for the controlled and reproducible fabrication of nanomembranes for multiple applications, such as photodetectors, field-effect transistors and ionizing radiation detectors for dosimetry.

**Acknowledgements**

The authors acknowledge the financial support from the Portuguese Foundation for Science and Technology (FCT) via the IonProGO project (2022.05329.PTDC, http://doi.org/10.54499/2022.05329.PTDC) and via the INESC MN Research Unit funding (UID/05367/2020) through Pluriannual BASE and PROGRAMATICO financing. The authors also acknowledge the high-performance computing resources provided by the Navigator platform of the Laboratory for Advanced Computing at University of Coimbra under the MDGaO FCT advanced computing projects (2023.09323.CPCA, https://doi.org/10.54499/2023.09323.CPCA.A1, and 2023.11548.CPCA, https://doi.org/10.54499/2023.11548.CPCA.A1), as well as the Finnish IT Center for Science (CSC). D. M. Esteves thanks FCT for his PhD grant (2022.09585.BD). The implantations were performed under proposal 22002965-ST of the RADIATE project (H2020: 824096, http://doi.org/10.3030/824096) and proposal 26001 of the ReMade@ARI project (https://doi.org/10.3030/101058414), funded by the European Union as part of the Horizon Europe call HORIZON-INFRA-2021-SERV-01 under grant agreement number 101058414 and co-funded by UK Research and Innovation (UKRI) under the UK government's Horizon Europe funding guarantee (grant number 10039728) and by the Swiss State Secretariat for Education, Research and Innovation (SERI) under contract number 22.00187. Views and opinions expressed are however those of the author(s) only and do not necessarily reflect those of the European Union or the UK Science and Technology Facilities Council or the Swiss State Secretariat for Education, Research and Innovation (SERI). Neither the European Union nor the granting authorities can be held responsible for them. The M-ERA.NET Program is acknowledged for financial support via the GOFIB project (administrated by the Research Council of Norway project number 337627 in Norway and the Academy of Finland project number 352518 in Finland). Additionally, the Research Council of Norway is acknowledged for support to the Norwegian Center for Transmission Electron Microscopy, NORTEM (No. 197405/F50), to the Norwegian Micro- and Nano-Fabrication Facility, NorFab (No. 295864), and DIOGO project (No. 351033).

# Supplementary Information

# Microtubes and nanomembranes by ion-beam-induced exfoliation of $\beta$-Ga$_2$O$_3$

D. M. Esteves[1,2,*], R. He[3], C. Bazioti[4], S. Magalhães[2,5], M. C. Sequeira[6], L. F. Santos[7], A. Azarov[4], A. Kuznetsov[4], F. Djurabekova[3], K. Lorenz[1,2,5], M. Peres[1,2,5]

**Graphical abstract:**

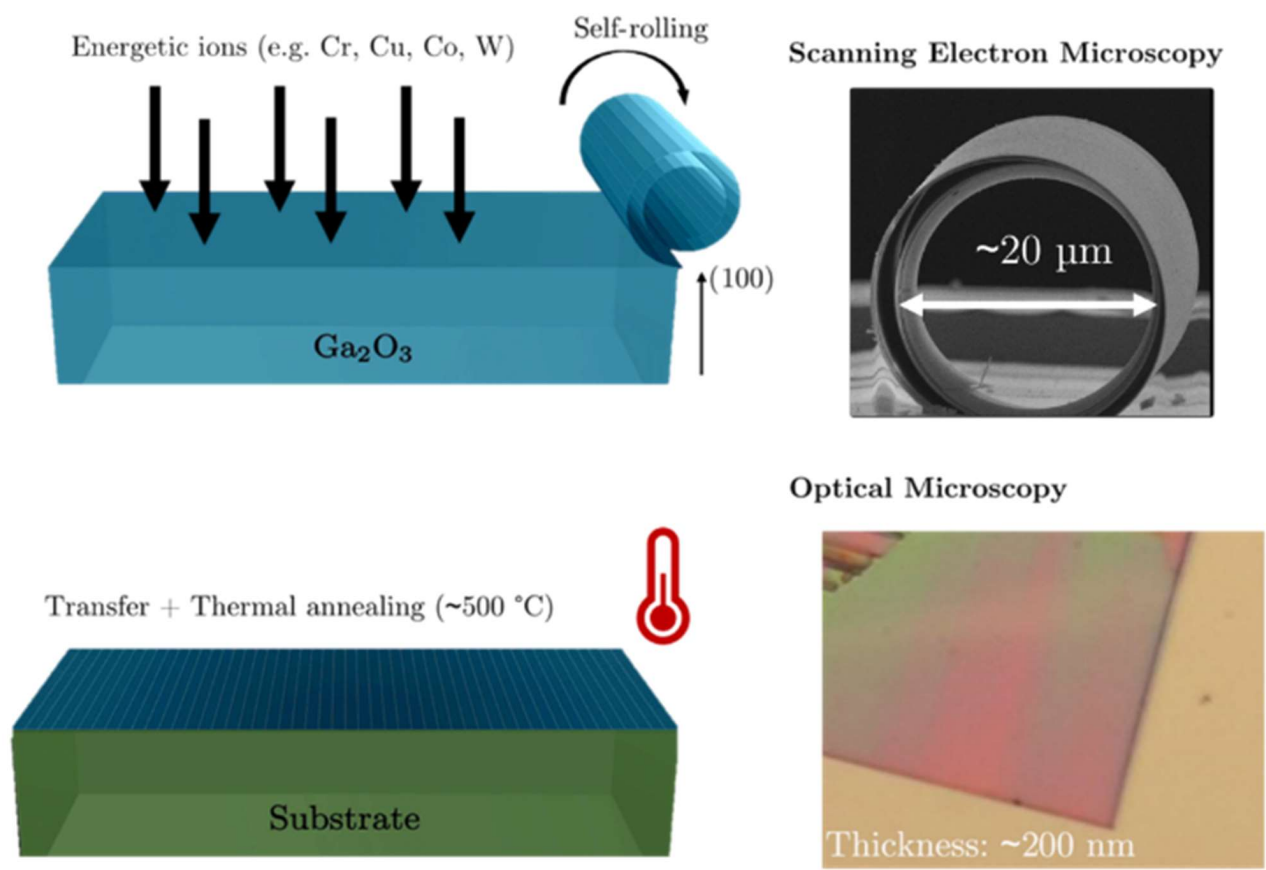

## 1. Critical fluence study

In order to pinpoint the threshold fluence for the formation of microtubes, a set of six nominally undoped commercial $\beta$-$Ga_2O_3$ single-crystals with a (100) surface orientation was implanted with 250 keV $Cr^+$ ions at IBC-HZDR with fluences between $6.0\times10^{12}$ and $6.0\times10^{14}$ cm$^{-2}$, at room temperature, with a constant flux of $2.0\times10^{10}$ cm$^{-2}$ s$^{-1}$. Fig. S1 shows optical microscopy images of the implanted samples to inspect the presence of microtubes.

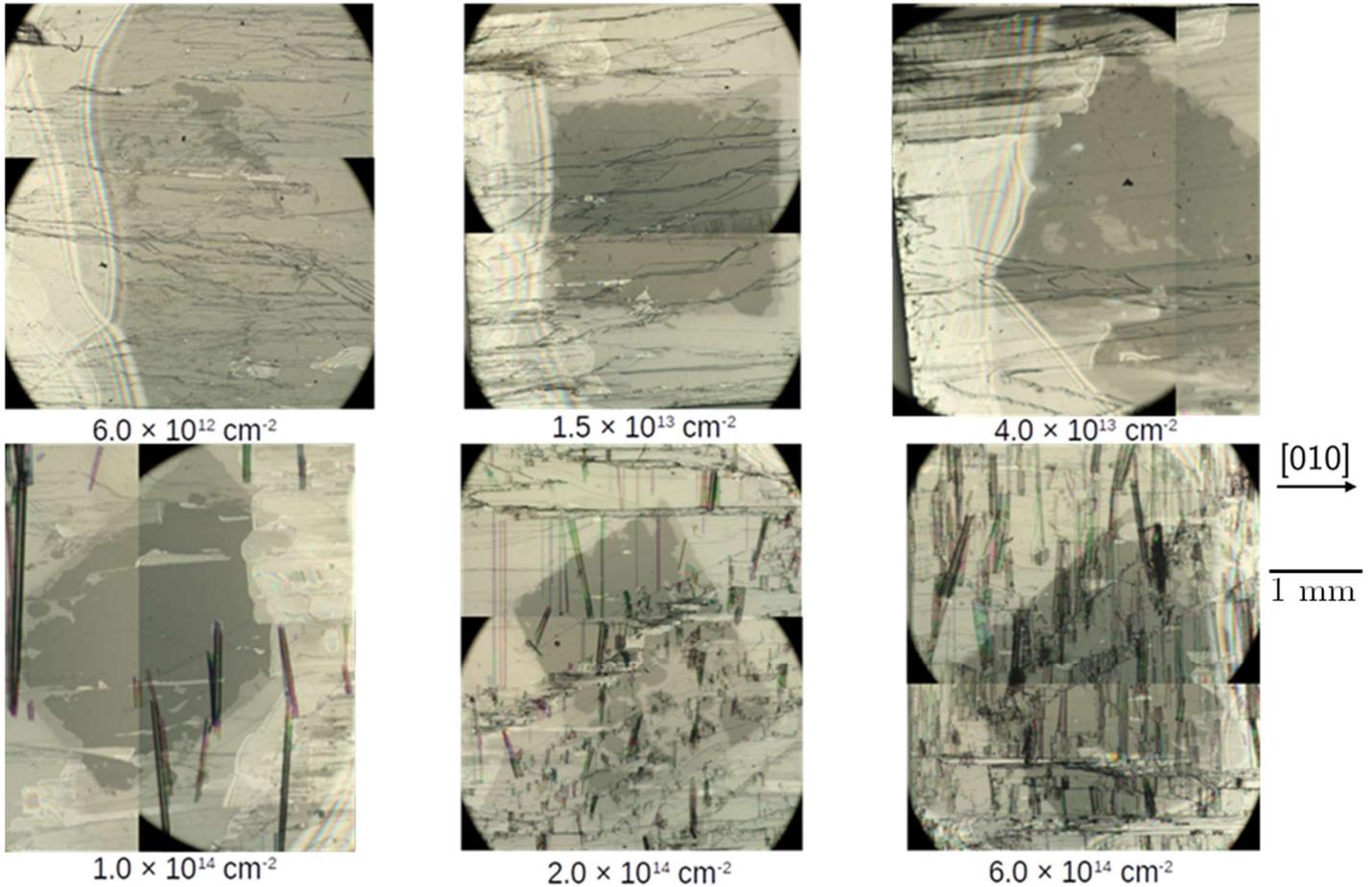

**Fig S1** | Optical microscopy image of the implanted samples to fluences between $6.0\times10^{12}$ and $6.0\times10^{14}$ cm$^{-2}$, under an implantation flux of $2.0\times10^{10}$ cm$^{-2}$ s$^{-1}$.

It is possible to observe that samples implanted with fluences up to $4.0\times10^{13}$ do not show the formation of microtubes, whereas those with fluences larger than $1.0\times10^{14}$ cm$^{-2}$ do show them on their surfaces. Hence, the critical fluence of the formation of microtubes should be of the order of ~$10^{14}$ cm$^{-2}$, under the considered implantation conditions. Regarding the size of the tubes and the yield of the process, it is possible to observe that, for larger fluence values, the number of tubes increases, while their size decreases. This observation may be related with the breakage of the fabricated tubes upon continuous implantation.

## 2. Membrane crystalline quality

The crystalline quality of the produced nanomembranes was assessed by μ-Raman spectroscopy and by High-Resolution Transmission Electron Microscopy (in the main paper). Fig S2 shows the Raman spectrum of a pristine bulk single-crystal sample (reference), as well the Raman spectra of membranes unrolled on top of a Si substrate by annealing at 500 and 1000 °C. The position and width of the peaks is very similar in all three samples, revealing that the crystalline quality of the nanomembranes is good and akin to a bulk single-crystal. Slight shifts and broadenings of the peaks after annealing at 500 °C are completely reversed at 1000 °C. Combined with the Transmission Electron Microscopy study in the main paper, these results suggest that the defect concentration at this dpa level is still small, and is quickly recovered by thermal annealing. These observations are in full agreement with the Rutherford Backscattering Spectrometry in Channeling mode and X-ray Diffraction measurements presented in the main paper.

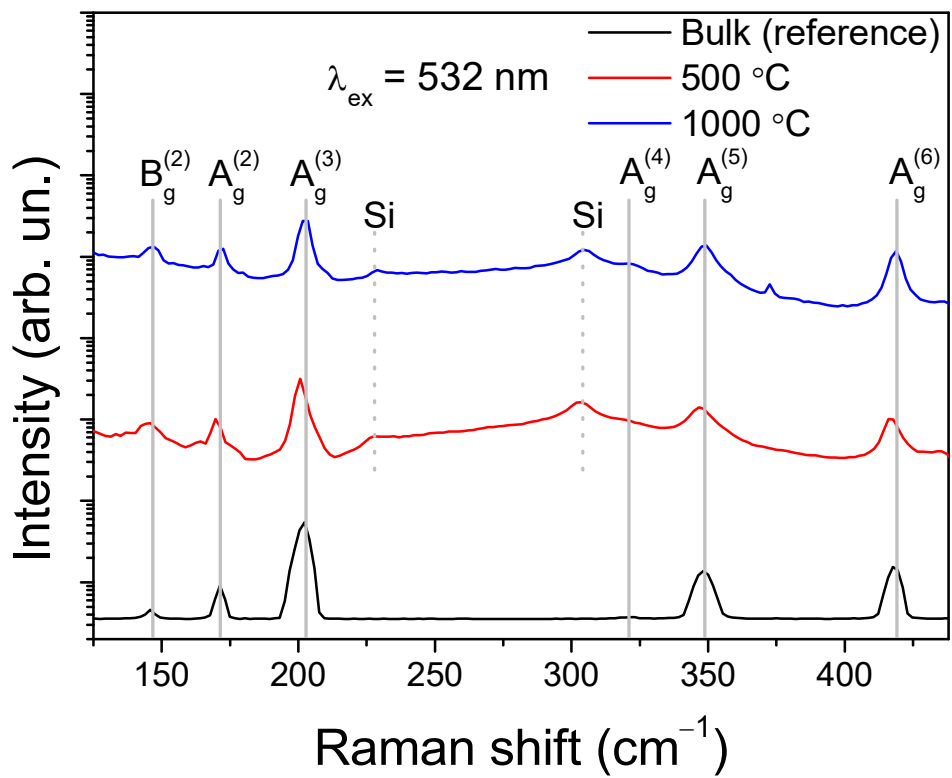

**Fig S2** | Raman spectra of a single-crystal bulk sample (reference) and of nanomembranes unrolled on top of a Si substrate after annealing at 500 and 1000 °C. The mode assignment follows the nomenclature proposed by T. Onuma et al. [1]

## 3. Bond method: high-accuracy lattice parameter determination

In order to obtain experimental reciprocal space maps in an absolute scale, it is necessary to measure the lattice parameters of the unimplanted crystal with high precision, thus fixing a point in the reciprocal space. In this work, Bond's method [2] was applied in order to measure the lattice parameters of the pristine sample. For this method, each reflection was measured twice: with grazing exit and with grazing incidence. The errors in the sample mounting, in particular those associated with the positioning of the surface of the sample, are thus cancelled, improving the accuracy.

The selected reflections were $600$ (symmetric), $710$ and $80\bar{1}$ (asymmetric). However, since the monoclinic system requires four lattice parameters ($a$, $b$, $c$ and $\beta$), a fourth reflection is necessary. The $002$ reflection was measured in skew-symmetric geometry, which is possible since the $(100)$ and $(001)$ planes are not perpendicular in this system. The Bond method results and lattice parameters are summarized in Table SI, where the uncertainties were propagated by adding in quadrature. The measured lattice parameters $a$, $b$, $c$, are slightly larger (by less than 0.2%) than those reported in the literature for single-crystals grown my a similar method [3], which may be related with the different experimental conditions or with the sample growth parameters.

| Experimental interplanar distances | |
|---|---|
| $d_{600} = \dfrac{1}{6} a \sin\beta$ | $(1.979 \pm 0.001)$ Å |
| $d_{710} = \dfrac{1}{\sqrt{\dfrac{1}{b^2} + \dfrac{49}{a^2 \sin\beta}}}$ | $(1.481 \pm 0.001)$ Å |
| $d_{80\bar{1}} = \dfrac{\sin\beta}{\sqrt{\dfrac{16\cos\beta}{ac} + \dfrac{64}{a^2} + \dfrac{1}{c^2}}}$ | $(1.529 \pm 0.001)$ Å |
| $d_{002} = \dfrac{1}{2} c \sin\beta$ | $(2.819 \pm 0.001)$ Å |
| **Calculated lattice parameters** | |
| $a = \dfrac{6\, d_{600}}{\sin\beta}$ | $(12.233 \pm 0.007)$ Å |
| $b = \dfrac{1}{\sqrt{\dfrac{1}{d_{710}^2} - \dfrac{49}{36\, d_{600}^2}}}$ | $(3.038 \pm 0.010)$ Å |
| $c = \dfrac{2\, d_{002}}{\sin\beta}$ | $(5.807 \pm 0.005)$ Å |
| $\beta = \arccos\left(\dfrac{3\, d_{600} d_{002}}{4\, d_{80\bar{1}}^2} - \dfrac{3\, d_{600}}{16\, d_{002}} - \dfrac{4\, d_{002}}{3\, d_{600}}\right)$ | $(103.882 \pm 0.174)°$ |

**Table SI** | Experimental interplanar distance and corresponding lattice parameters obtained from Bond's method of the virgin sample.

## 4. Reciprocal space relations in monoclinic systems

In order to better understand how the ion implantation impacted the three lattice constants $a$, $b$ and $c$, as well as the monoclinic angle $\beta$, the position of the peaks identified by the arrows in Fig. 3 of the main paper were used to calculate the strains and relative variation of the lattice parameters. The results are summarised in Table SII, where the subscript 0 pertains to the relaxed peak on the RSM. Since this analysis provides four values (the parallel and perpendicular components of the scattering vector for the two asymmetric reflections), it is possible to univocally determine $a$, $b$, $c$ and $\beta$. Moreover, it allows one to resolve the variation of the "out-of-plane lattice parameter" (here given by $x = a \sin \beta$) into two contributions: one due to the increase of the $a$ lattice parameter and another due to the decrease of the $\beta$ angle. Note that these strain values are in excellent agreement with those obtained in Fig 2. b) in the main paper.

|  | Reflection | |
|---|---|---|
|  | 710 | $80\bar{1}$ |
| $Q_{\perp 0}$ (nm$^{-1}$) | 37.035 | 39.651 |
| $Q_{\perp}$ (nm$^{-1}$) | 36.888 | 39.512 |
| $Q_{\|0} = Q_{\|}$ (nm$^{-1}$) | 20.679 | −10.820 |
| In-plane lattice parameters $b$ and $c$ (nm) | $b = b_0 = \dfrac{2\pi}{Q_\|^{710}} = 0.304$ | $c = c_0 = -\dfrac{2\pi}{Q_\|^{80\bar{1}}} = 0.581$ |
| "Out-of-plane lattice parameter" (nm) | $x = a \sin \beta = \dfrac{14\pi}{Q_\perp^{710}} = 1.192$ <br> $x_0 = a_0 \sin \beta_0 = \dfrac{14\pi}{Q_{\perp 0}^{710}} = 1.188$ | Strain: <br> $\varepsilon_\perp = \dfrac{Q_{\perp 0}^{710} - Q_\perp^{710}}{Q_\perp^{710}} = 0.40\%$ |
| Monoclinic angle (°) | $\beta = \arctan\left(\dfrac{Q_\|^{80\bar{1}}}{Q_\perp^{80\bar{1}} - 16\pi/x}\right) = 103.74$ <br> $\beta_0 = \arctan\left(\dfrac{Q_\|^{80\bar{1}}}{Q_{\perp 0}^{80\bar{1}} - 16\pi/x_0}\right) = 103.89$ | Relative variation: <br> $\varepsilon_\beta = \dfrac{\beta - \beta_0}{\beta_0} = -0.14\%$ |
| $a$ lattice parameter (nm) | $a = \dfrac{x}{\sin \beta} = 1.227$ <br> $a_0 = \dfrac{x_0}{\sin \beta_0} = 1.223$ | Strain: <br> $\varepsilon_a = \dfrac{a - a_0}{a_0} = 0.34\%$ |

**Table SII** | Experimental positions of the peaks in the RSM, as well as the calculated lattice parameters and associated strains or relative variations.

## 5. Molecular Dynamics simulations: additional details

While Molecular Dynamics is a very powerful simulation method, one of its weaknesses is the small scale of the simulation cells, due to the limitations imposed by the available computational power. As such, it is not possible to simulate the collision cascades created by 250 keV Cr ions. Therefore, the simulations in the main paper were performed using 10 keV Ga ions, which, as shown in Fig. S3, have similar maxima on the vacancies per ion and unit length profiles, according to Stopping and Ranges of Ions in Matter (SRIM) Monte Carlo simulations [4]. Under these conditions, the dpa value at the maximum of the damage curve is similar in the two conditions, which facilitates the direct comparison between experiment and simulation. Moreover, it is quite convenient to use Ga ions as the projectile, as it is already contained in the interaction potential.

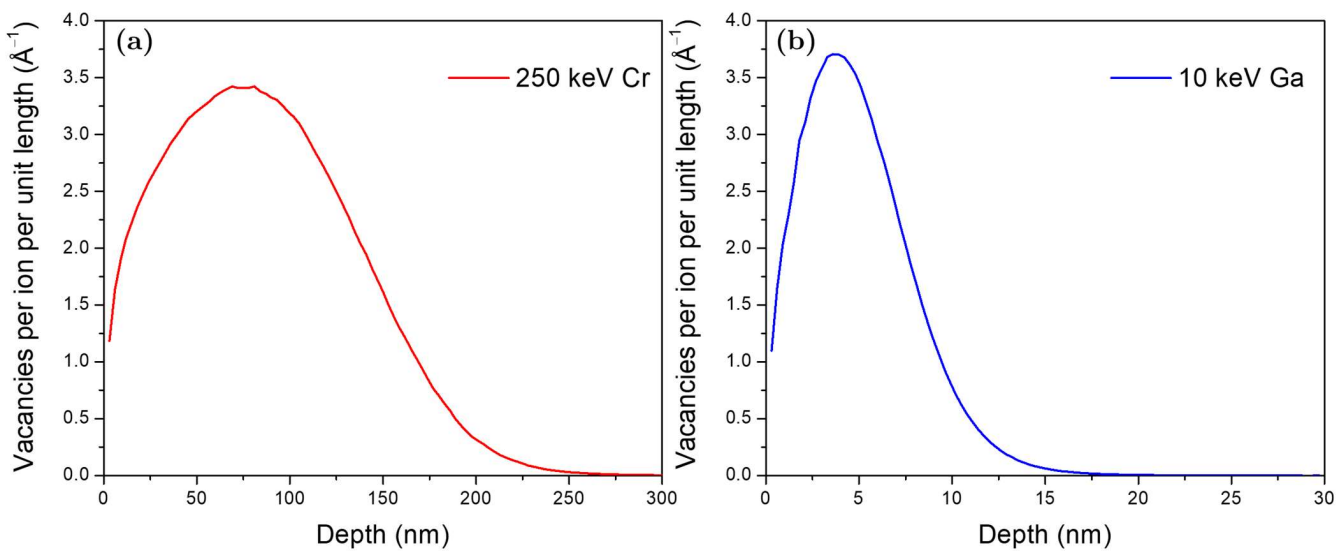

**Fig. S3** | SRIM Monte Carlo simulations comparing the vacancies per ion and per unit length profiles in the case of 250 keV Cr (a) or 10 keV Ga ions (b).

Regarding the simulation details given in the experimental section of the main paper, three regions were defined inside the simulation cell, as shown in Fig. S4. The main part of the cell consisted of a layer extending from the surface (shown in the left-hand side of Fig. S4) down to a depth of ~220 Å, following the microcanonical (NVE) ensemble, within which the collision cascade developed fully. This was followed by a 20-Å thermostat layer employing the NVT ensemble, which was modelled with a Nosé-Hoover thermostat at 300 K [5]. This layer aims to simulate the energy dissipation by the substrate, thus removing energy from the NVE region above. Finally, a 10-Å boundary layer of fixed atoms was employed so as to fix the position of the whole cell.

The projectiles consisted of Ga atoms that were initialized at random positions in the plane 20 Å away from the surface (in order to avoid direct interaction with the surface atoms), with a kinetic energy of 10 keV. Additionally, the channeling effect was suppressed by setting an angle of 7° with respect to the direction perpendicular to the (100) plane, which corresponds approximately to the [201] direction.

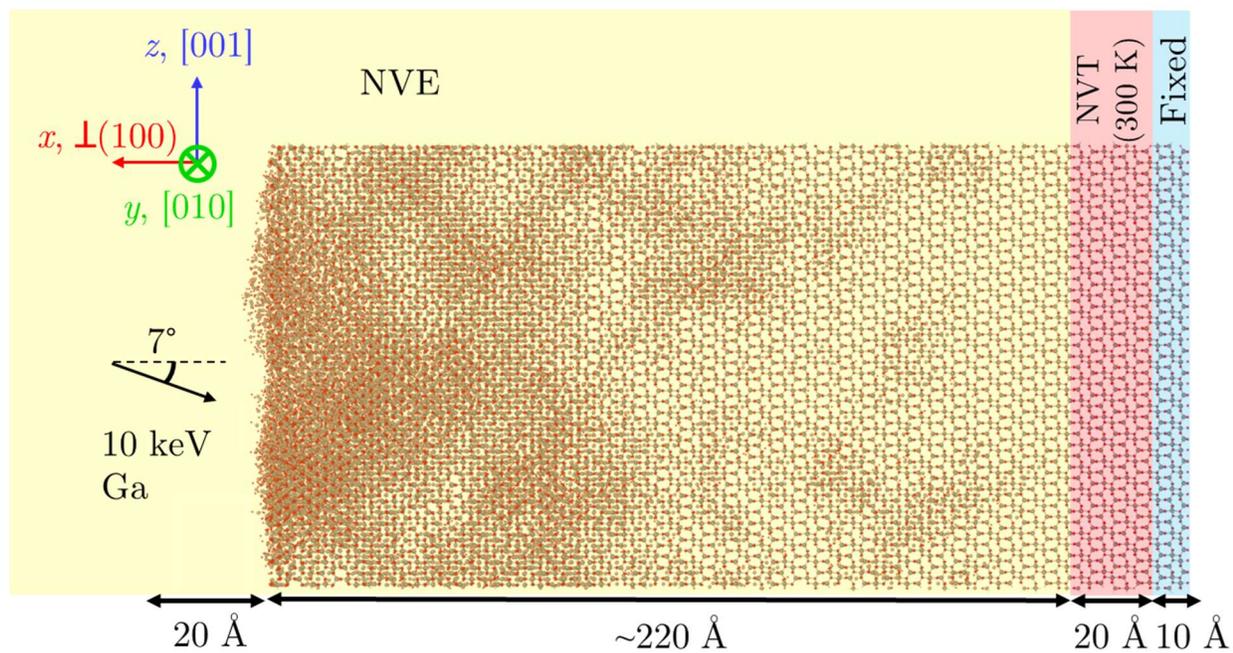

**Fig. S4** | Regions and associated ensembles defined on the simulation cell for the implantation simulation.

## 6. Simulated reciprocal space maps

Combining the values of $Q_\parallel$ and $Q_\perp$ for each of the two simulated reciprocal space maps of the asymmetric reflections shown in Fig. 9 in the main paper, it is possible to determine the lattice parameters ($a$, $b$, $c$, $\beta$) and the associated strain/relative variations, as indicated in table SIII.

The lattice parameters of the virgin sample are in close agreement to those previously reported using this interatomic potential [6] (1.2480 nm, 0.3086 nm, 0.5871 nm and 103.880° for $a$, $b$, $c$ and $\beta$, respectively), and the small variations are attributed to the different relaxation conditions imposed by the layer structure of Fig. S3. It is important to note that the $Q_\perp$ coordinate of $80\bar{1}$ involves both $a$ and $c$, due to the non-orthogonality between the $a$- and $c$-axes, whereas the $Q_\parallel$ and $Q_\perp$ coordinates of $710$ involve only $b$ or $a$, since the corresponding axes are orthogonal. The expressions for the reciprocal space quantities were derived using a Mathematica script [7] employing the metric tensor of the monoclinic system in order to compute the interplanar distances and angles between the desired planes [8].

| Parameter | Virgin | Damaged | Strain |
|---|---|---|---|
| $Q_\parallel^{710} = \dfrac{2\pi}{b}$ | 20.32 nm$^{-1}$ | 20.32 nm$^{-1}$ | — |
| $Q_\perp^{710} = \dfrac{14\pi}{a \sin\beta}$ | 36.187 nm$^{-1}$ | 36.067 nm$^{-1}$ | — |
| $Q_\parallel^{80\bar{1}} = -\dfrac{2\pi}{c}$ | −10.68 nm$^{-1}$ | −10.68 nm$^{-1}$ | — |
| $Q_\perp^{80\bar{1}} = 2\pi\left(\dfrac{8}{a\sin\beta} + \dfrac{1}{c\tan\beta}\right)$ | 38.706 nm$^{-1}$ | 38.625 nm$^{-1}$ | — |
| $a = \dfrac{14\pi}{Q_\perp^{710}}\sqrt{1 + \left(\dfrac{Q_\perp^{710} - Q_\perp^{80\bar{1}}}{Q_\parallel^{80\bar{1}}}\right)^2}$ | 1.249 nm | 1.254 nm | 0.40% |
| $b = \dfrac{2\pi}{Q_\parallel^{710}}$ | 0.3092 nm | 0.3092 nm | 0% |
| $c = -\dfrac{2\pi}{Q_\parallel^{80\bar{1}}}$ | 0.5883 nm | 0.5883 nm | 0% |
| $\beta = \mathrm{arccot}\left(\dfrac{\frac{8}{7}Q_\perp^{710} - Q_\perp^{80\bar{1}}}{Q_\parallel^{80\bar{1}}}\right)$ | 103.94° | 103.65° | −0.28%* |
| $d_{600} = \dfrac{a\sin\beta}{6}$ | 0.2020 nm | 0.2031 nm | 0.54% |

**Table SIII** | Positions of the maxima of the simulated RSM after 20 overlaps (~0.05 dpa) and the corresponding lattice parameters calculated from them. The asterisk (*) denotes that the relative variation of $\beta$ is not really a strain.